\documentclass[aps,floatfix,pre,notitlepage]{revtex4-1}
\usepackage{latexsym,amssymb,graphicx,amsmath,epsfig,calc,times}
\usepackage{chemarr,bbding,wasysym}

\usepackage{graphicx}
\usepackage[usenames]{color}
\usepackage{latexsym,amssymb,amsmath,epsfig,calc}
\usepackage{chemarr,bbding,wasysym}

\newcommand{\Gcap}{G^\text{cap}}
\newcommand{\Fec}{F_\text{EC}}
\newcommand{\ncontact}{n^\text{c}}

\newcommand{\sdegen}{S^{\text{degen}}}
\newcommand{\gb}{g_{\text{b}}}
\newcommand{\eb}{\varepsilon_{\text{b}}}
\newcommand{\gc}{g_{\text{s}}}
\def\gnuc{g_\text{nuc}}
\def\gelong{g_\text{elong}}
\def\nnuc{{n_\text{nuc}}}
\def\tnuc{\tau_\text{nuc}}
\def\nelong{n_\text{elong}}
\def\telong{\tau_\text{elong}}
\newcommand{\halftime}{\tau_\text{1/2}}
\newcommand{\nn}{{\nnuc-1}}
\newcommand{\kt}{k_{\text{B}}T}
\def\fc{f_\text{c}}
\def\fp{f_\text{p}}
\def\fceq{f_\text{c}^\text{eq}}
\def\rhoTot{\rho_\text{T}}
\def\rhop{\rho_\text{p}}
\def\rhopt{\rho_\text{pT}}
\def\coreTot{x_\text{T}}
\def\phiTot{\phi_\text{T}}

\def\rhoStar{\rho^*}
\def\rhoStarStar{\rho^{**}}
\def\cc{\rho_\text{c}}
\def\ckt{\rho_\text{kt}}

\def\stoich{r}
\def\Lp{N_\text{p}}
\def\lp{N_\text{p}}
\def\ecp{\varepsilon_\text{cp}}
\def\ecc{\varepsilon_\text{cc}}

\newcommand{\logc}{\log\rhoTot}

\newcommand{\CC}{\rho_\text{C}}

\newcommand{\Gcore}{G^\text{core}}

\definecolor{Blue}{rgb}{0,0.0,1.0}
\definecolor{Red}{rgb}{1.0,0.0,0.0}
\definecolor{Green}{rgb}{0.0,0.35,0.0}
\definecolor{Grey}{rgb}{0.5,0.5,0.5}

\begin{document}

\title{Modeling Viral Capsid Assembly}
\author{Michael F. Hagan}
\affiliation{Department of Physics, Brandeis University, Waltham, MA, 02454}
\email{hagan@brandeis.edu}
\date{\today}

\begin{abstract}
I present a review of the theoretical and computational methodologies that have been used to model the assembly of viral capsids. I discuss the capabilities and limitations of approaches ranging from equilibrium continuum theories to molecular dynamics simulations, and I give an overview of some of the important conclusions about virus assembly that have resulted from these modeling efforts. Topics include the assembly of empty viral shells, assembly around single-stranded nucleic acids to form viral particles, and assembly around synthetic polymers or charged nanoparticles for nanotechnology or biomedical applications. I present some examples in which modeling efforts have promoted experimental breakthroughs, as well as directions in which the connection between modeling and experiment can be strengthened.
\end{abstract}

%\contentsinbrief %optional

\maketitle

\tableofcontents

\section{Introduction}
\label{sec:intro}
The formation of a virus is a marvel of natural selection.  A large number (~60-10,000) of protein subunits and other components assemble into complete, reproducible structures, often with extreme fidelity, on a biologically relevant time scale. Viruses play a role in a significant portion of human diseases, as well as those of other animals, plants, and bacteria. Thus, it is of great interest to understand their formation process, with the goal of developing novel antivirus therapies that can block it, or alternatively to re-engineer viruses as drug delivery vehicles that can assemble around their cargo and disassemble to deliver it without requiring explicit external control. More fundamentally, learning the factors that make viral assembly so robust could advance the development of self-assembling nanostructured materials.

This review focuses on the use of theoretical and computational modeling to understand the viral assembly process. We begin with brief overviews of virus structure, assembly, and the experiments used to characterize the assembly process (section \ref{sec:intro}). We next perform an equilibrium analysis of the assembly of empty protein shells in section \ref{sec:thermoSection}. In section \ref{sec:dynamics} we then present a simple mathematical representation of the assembly process and its relevant timescales, followed by several types of modeling approaches that have been used to  analyze and predict \emph{in vitro} assembly kinetics. We then extend the equilibrium and dynamical approaches to consider the co-assembly of capsid proteins with RNA or other polyanionic cargoes in section \ref{sec:cargo}. Finally, we conclude with  some of the important  open questions and ways in which modeling can  make a stronger connection with experiments.

In the interests of thoroughly examining the capsid assembly process, this review will not consider a number of interesting topics such as the structural dynamics of complete capsids (e.g. \cite{Freddolino2006,Arkhipov2006,Joshi2011a,Joshi2011}), capsid swelling or maturation transitions (e.g. \cite{Guerin2007,Miao2010,May2011,Tama2005,May2012,Roos2012,May2012a,Widom2007,May2011a}), mechanical probing of assembled capsids (e.g. \cite{Roos2012,Arkhipov2009,Klug2006,Baclayon2010,Gibbons2008,Gibbons2007a,Michel2006,Roos2009,Gibbons2007,Roos2010,Roos2010a,Cieplak2010}),
or the motor-driven packaging of double-stranded DNA (dsDNA) into assembled procapsids (\cite{Locker2006,Forrey2006,Petrov2009,Petrov2008,Petrov2007a,Petrov2007,Rollins2008}, reviewed in \cite{Harvey2009,Angelescu2008b,Siber2012}), or the conformations of dsDNA inside capsids (e.g. \cite{Leforestier2009,Leforestier2010,Leforestier2011}).

%Other reviews:
%\begin{itemize}
%\item Zlotnick 2005, Spier, other COCB (Harrison).
%\item Podgornik (reference this also in electrostatics area) \cite{Siber2012}
%\item \cite{Bruinsma2006}
%\item Linse COCB, Soft Matter
%\item Harvey2009
%\end{itemize}

\subsection{Virus anatomies}
\label{sec:anatomies}
Viruses consist of at least two types of components: the genome, which can be DNA or RNA and can be single-stranded or double-stranded, and a protein shell called a capsid that surrounds and protects the fragile nucleic acid.  Viruses vary widely in complexity, ranging from satellite tobacco mosaic virus (STMV), whose 1063-nucleotide genome encodes for two proteins including the capsid protein \cite{Routh1995} to the giant Megavirus, with a 1,259,197-bp genome encoding for 1,120 putative proteins \cite{Arslan2011} that is larger than some bacterial genomes and encased in two capsids and a lipid bilayer.  Viruses such as Megavirus that acquire a lipid bilayer coating from the plasma membrane or  an interior cell compartment of the host organism are known as `enveloped' viruses, whereas viruses such as STMV that present a naked protein exterior are termed `non-enveloped'. Since Stephen Harrison and colleagues achieved the first atomic-resolution structure of tomato bushy stunt virus (TBSV) in 1978 \cite{Harrison1978}, structures of numerous virus capsids  have been revealed to atomic resolution by x-ray crystallography and/or cryo-electron microscopy (cryo-EM) images. An extensive collection of virus structures can be found at the VIPERdb virus particle explorer website (http://viperdb.scripps.edu) \cite{Reddy2001}.

The requirement that the viral genome be enclosed in a protective shell severely constrains its length and hence the number of protein sequences that it can encode. As first proposed by Crick and Watson \cite{Crick1956}, virus capsids are therefore comprised of numerous copies of one or a few protein sequences, which are usually arranged with a high degree of symmetry in the assembled capsid. Most viruses can be classified as rodlike or spherical, with the capsids of rodlike viruses arranged with helical symmetry around the nucleic acid, such as tobacco mosaic virus (TMV), and the capsids of most spherical viruses arranged with icosahedral symmetry. There are also important exceptions discussed below. The number of protein copies comprising a helical capsid is arbitrary and thus a helical capsid can accommodate a nucleic acid of any length. In contrast, icosahedral capsids are limited by the geometric constraint that at most 60 identical subunits can be arranged into a regular polyhedron. However, early structural experiments indicated that many spherical capsids contain multiples of 60 proteins.

\begin{figure}
\begin{center}
\epsfxsize=0.6\textwidth\epsfbox{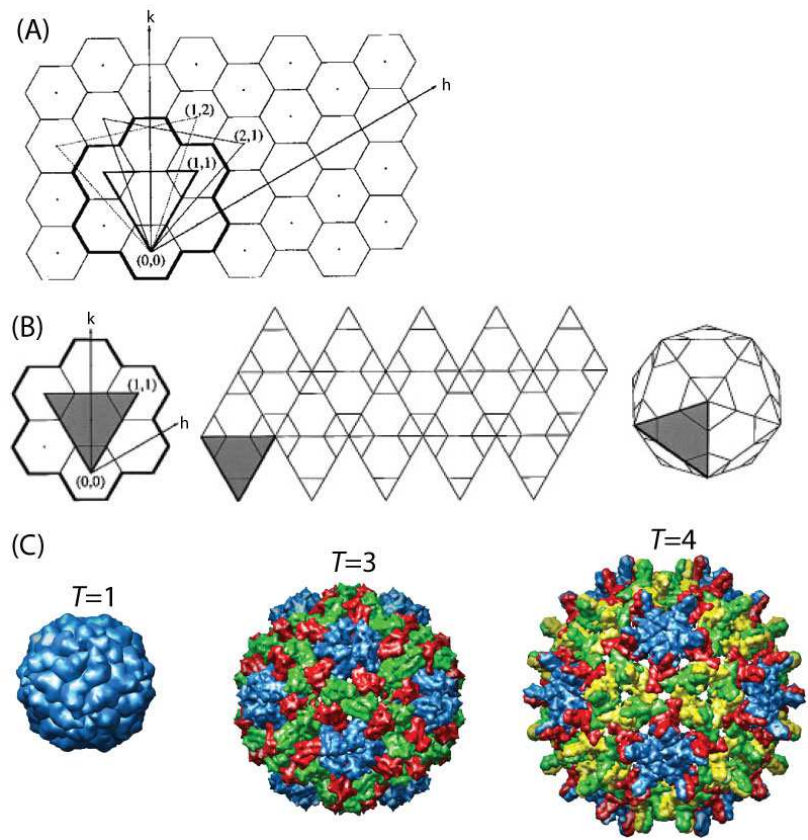}
\caption{  The geometry of icosahedral lattices. {\bf (A)} Different equilateral triangular facets can be constructed on a hexagonal lattice by moving integer numbers of steps along each of the $\mathbf{\hat{h}}$ and $\mathbf{\hat{k}}$ lattice vectors.  {\bf (B)}  Construction of a $T$=3 lattice. Twenty copies of the triangular facet (left) obtained by moving one step along each of the $\mathbf{\hat{h}}$ and $\mathbf{\hat{k}}$ lattice vectors are arranged as shown in the middle panel, and then folded to obtain the icosahedral structure shown on the right.  To connect this construction to a capsid, note that each pentagon will comprise five proteins in identical environments and each hexagon will comprise six subunits in two different types of local environments, resulting in a total of 180 proteins in three distinct local environments.  {\bf (C)} Example icosahedral capsid structures. From left to right are shown the $T$=1 satellite tobacco mosaic virus capsid (STMV) PDBID 1A34 \cite{Larson1998}, the $T$=3 cowpea chlorotic mottle virus capsid (CCMV) PDBID 1CWP \cite{Speir1995}, and the $T$=4 human hepatitis B viral capsid (HBV) PDBID 1QGT \cite{Wynne1999}. Structures are shown scaled to actual size, and the protein conformations are indicated by color. In each image the 60 pentameric subunits are colored blue. The images of capsids in (C) were obtained from the Viper database \cite{Reddy2001}.  The images in (A) and (B) were reprinted from J. Mol. Biol., Johnson and Speir, {\bf 269}, 665-675 (1997) \emph{Quasi-equivalent viruses: A paradigm for protein assemblies}, with permission from Elsevier.
\label{fig:tNumber}
}
\end{center}
\end{figure}

  Caspar and Klug proposed geometrical arguments that describe how multiples of 60 proteins can be arranged with icosahedral symmetry, where individual proteins interact through the same interfaces but take slightly different, or quasi-equivalent, conformations \cite{Caspar1962}.  Protein subunits can be grouped into morphological units or `capsomers', usually as pentamers and hexamers.  Icosahedral symmetry requires exactly 12 pentamers, located at the vertices of an icosahedron inscribed within the capsid.  A complete capsid is comprised of $60T$ subunits, where $T$ is the `triangulation number', which is equal to the number of distinct subunit conformations.

In brief, a structure with icosahedral symmetry is comprised of 20 identical facets. The facets are equilateral triangles and thus themselves comprise at least 3 identical asymmetric units (asu). The only requirement of the asymmetric units  is that they are arranged with threefold symmetry, although many capsid proteins have a roughly trapezoidal shape \cite{Rossmann1989} and it has been argued that this shape is ideal for tiling icosahedral structures \cite{Mannige2008}. The Caspar Klug (C-K) classification system can be obtained starting from a hexagonal lattice as shown in Fig.~\ref{fig:tNumber}. An edge of the icosahedral facet is defined by starting at the origin and stepping distances $h$ and $k$ along each of the respective lattice vectors. There is an infinite series of such equilateral triangles corresponding to integer values of $h$ and $k$. The area of such a triangle (for unit spacing between lattice points) is given by $T/4$, where $T$ is the triangulation number defined as
\begin{equation}
T=h^2+hk+k^2
\label{eq:tNumber}.
\end{equation}
Considering that the smallest such triangle $T$=1 comprises three asu, the total number of asu in the facet is thus given by $3T$ and the total number of asu in the icosahedron is $60T$. From Fig.~\ref{fig:tNumber} the individual asu's are not all identical for $T>1$ since they have different local environments.  Given the threefold symmetry of the facet, there are $T$ distinct local environments and thus $T$ distinct asu geometries.  Fig.~\ref{fig:tNumber}B shows how to build a physical model for such a construction with $T$=3.

The asu (i.e. individual proteins) within the icosahedral structure can be grouped based on whether they sit at a five-fold or threefold (quasi-sixfold) axis of symmetry into pentameric or hexameric `capsomers'. Given that an icosahedron contains 12 vertices with fivefold symmetry and the total number of proteins is given by $60T$, there are $10(T-1)$ hexamers.

Many icosahedral viral capsids with $T>1$ are  comprised of only a single protein copy, meaning that the protein must adopt different configurations depending on its local environment. It was originally proposed by Caspar and Klug \cite{Caspar1962} that because the local environments are similar, or `quasi-equivalent', the proteins in different environments could interact through the same interfaces. This has since been found to be correct for many icosahedral viruses, with structural differences between proteins at different quasi-equivalent sites often limited to loops and N- and C-termini. However, there can also be proteins with extensive conformational changes or even different sequences at different sites. Some examples of these structural differences are reviewed in Refs.\cite{Johnson1997,Zlotnick2005}.

 Some icosahedral virus capsid structures deviate from the class of lattice structures shown in  Fig.~\ref{fig:tNumber}. For example, the {\em Polyomaviridae}, e.g. human papilloma virus (HPV), are comprised entirely of pentamer capsomers, which depending on their local environment are either five-fold or six-fold coordinated. Generalizations of the C-K classification scheme have been proposed \cite{Twarock2004,Keef2008,Keef2009,Keef2008a,Lorman2007,Konevtsova2012,Lorman2008} which can describe polyomavirus capsid shapes.  Mannige and Brooks identified a relationship between hexamer shapes and capsid properties such  as size \cite{Mannige2008,Mannige2009}. They also developed a metric for complexity of icosahedral morphologies, which resulted in a `periodic table' of capsids and, combined with the assumption that the simplest structures are the fittest, revealed evolutionary pressures on capsid structures \cite{Mannige2010}.

    There are also non-spherical capsids with aspects of icosahedral symmetry.  For example, the mature HIV virus capsid assembles into tubular or conical shapes
    \cite{Ganser-Pornillos2004,Benjamin2005,Ganser-Pornillos2007,Ganser-Pornillos2008} and some bacteriophages (viruses that infect bacteria) have capsids which are elongated or prolate icosahedra (e.g. \cite{Moody1965,Fokine2004}). The C-K classification system was extended to describe prolate icosahedra by Moody \cite{Moody1965}. We present some approaches to model the stability and formation of capsids that correspond to C-K structures or their generalizations in section \ref{sec:higherTNumber}.

\subsection{Virus assembly}
\label{sec:introVirusAssembly}
Viral assembly most generally refers to the process by which the protein capsid(s) form, the nucleic acid becomes encapsulated within the capsid, membrane coats are acquired (if the virus is enveloped), and any maturation steps occur. For many viruses the capsid can form spontaneously, as demonstrated in 1955 by the experiments of Fraenkel-Conrat and Williams in which the RNA and capsid protein of TMV spontaneously assembled \emph{in vitro} to form infectious virions \cite{Fraenkel-Conrat1955}.

The pathway of nucleic acid encapsulation differs dramatically between viruses with single-stranded or double-stranded genomes. Viruses with  single-stranded genomes (the best studied of which have ssRNA genomes) usually assemble spontaneously around their nucleic acid in a single step.  This category includes many small spherical plant viruses (e.g. STMV or {\em bromoviridae}), the bacteriophage MS2, and animal viruses such as nodavirus. In many cases the RNA is required for assembly at physiological conditions, but the capsid proteins can assemble without RNA into empty shells \emph{in vitro} under different ionic strengths or $p$H. We also can include in this group the \emph{Hepadnaviridae} family of viruses (e.g. Hepatitis B Virus (HBV)), which have a dsDNA genome but a capsid that assembles around an ssRNA pregenome \cite{Kann1994,Bartenschlager1992,Hu1997}.

The extreme stiffness of a double-stranded genome (the persistence length of dsDNA is 50 nm) and the high charge density preclude spontaneous nucleic acid encapsidation. Thus packaging a double-stranded genome requires a two-step process in which an empty protein shell is assembled followed by packaging via ATP hydrolysis and/or complexation with nucleic acid folding proteins (e.g. histones \cite{Griffith1975,Varshavsky1976}). Of these viruses, the assembly processes have been most thoroughly investigated for dsDNA viruses, such as the tailed bacteriophages, herpes virus and adenovirus. These viruses assemble an empty capsid, without requiring a nucleic acid at physiological conditions, and a molecular motor which inserts into one vertex of the capsid \cite{Sun2010}. The motor then hydrolyzes cellular ATP to pump the DNA into the capsid.

In this review we will focus on the assembly of icosahedral viruses, first discussing the assembly of empty capsids such as occurs during the first step of bacteriophage assembly, and then  co-assembly of capsid proteins with RNA, such as occurs during replication of ssRNA viruses, and finally co-assembly with other polyanions  in \emph{in vitro} experiments.  We will not consider the assembly of rod-like viruses. Although not yet completely understood, the assembly process for the rod-like virus TMV has been studied in great detail and has been the subject of several reviews \cite{Butler1999,Caspar1990,Klug1999} as well as more recent modeling studies \cite{Kegel2006,Kraft2012}.

\subsubsection{Experiments that characterize capsid assembly}
\label{sec:introExperiments}
\begin{figure}
\begin{center}
\epsfxsize=0.4\textwidth\epsfbox{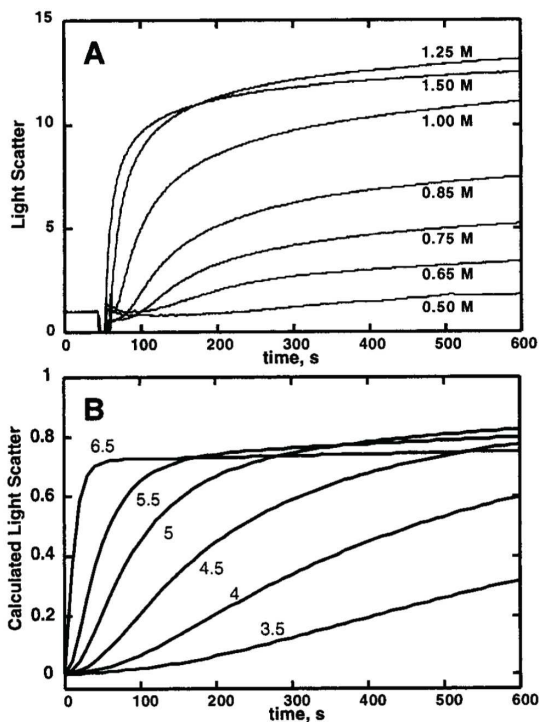}
\caption{  {\bf (A)} Light scattering  measured as a function of time for 5 $\mu$M dimer of HBV capsid protein at indicated ionic strengths.  Light scatter is approximately proportional to the mass-averaged molecular weight of assemblages and, under conditions of productive assembly, closely tracks the fraction of subunits in capsids (see text). {\bf (B)} Simulated light scattering for 5 $\mu$M subunit with indicated values of the subunit-subunit binding free energy ($\gb$) using the rate equation approach described in section \ref{sec:rateEquations}. Figures reprinted
with permission from Biochemistry, {\bf 38}, 14644-14652 (1999), \emph{A Theoretical Model Successfully Identifies Features of Hepatitis B Virus Capsid Assembly}, Zlotnick, Johnson, Wingfield, Stahl, Endres,  Copyright (1999) American
Chemical Society.}
 \label{fig:lightScatter}
\end{center}
\end{figure}

The kinetics for spontaneous capsid assembly \emph{in vitro} have been measured with size exclusion chromatography (SEC) and X-ray and light scattering (e.g. \cite{Prevelige1993,Zlotnick1999,Zlotnick2000,Casini2004,Johnson2005,Chen2008,Berthet-Colominas1987,Kler2012}). Most frequently, the fraction of subunits in capsids or other intermediates has been monitored using size exclusion chromatography (SEC) and the mass-averaged molecular weight has been estimated with light scattering.
The SEC experiments show that under optimal assembly conditions the only species present in detectable concentrations are either complete capsids or small oligomers which we refer to as the basic assembly unit. The size of the basic assembly unit is virus dependent; e.g. dimers for bromoviruses \cite{Adolph1974} and HBV \cite{Ceres2002,Wingfield1995}, or pentamers for picornaviruses (e.g. human rhinovirus (HRV)) and the \emph{polyomaviridae} family \cite{Li2002} (e.g. human papilloma virus (HPV)).
Provided that intermediate concentrations remain small, the mass-averaged molecular weight and thus the light scattering closely track the fraction capsid measured by SEC. Example light scattering measurements from Zlotnick and coworkers \cite{Zlotnick1999} are shown in Fig. \ref{fig:lightScatter}A for HBV assembly at several ionic strengths.

While these bulk experiments have provided tremendous information about capsid assembly kinetics, it has been difficult to characterize assembly pathways in detail because the intermediates are transient and present only at low levels. Complementary techniques have begun to address this limitation. Restive pulse sensing was used to track the passage of individual HBV capsids through conical nanopores in a membrane \cite{Zhou2011,Harms2011}.
This  Coulter-counter-like apparatus was able to distinguish between $T$=3 and $T$=4 capsids.  Mass spectrometry has been used to characterize key intermediates in the assembly of MS2 by Stockley and coworkers \cite{Stockley2007,Toropova2008,Basnak2010} (see section \ref{sec:largeTDynamics}) and for HBV and nodavirus by Uetrecht et al. \cite{Uetrecht2011}.
Furthermore, fluorescent labeling of capsid proteins \cite{Jouvenet2008} and in some cases RNA has enabled measuring assembly timescales for capsids \emph{in vivo} (reviewed in Refs. \cite{Baumgaertel2012,Jouvenet2011}).

\subsubsection{Motivation for and scope of modeling}
\label{sec:introMotivaton}
Even with the experimental  capabilities to detect and characterize key intermediates for some viruses, theoretical and computational models are important complements to elucidate assembly pathways and mechanisms. Each intermediate is a member of a large ensemble of structures and pathways that comprise the overall assembly process for a virus.   Furthermore, assembly is driven by collective interactions that are regulated by a tightly balanced competition of forces between individual molecules.  It is difficult, with experiments alone, to parse these interactions for those mechanisms and factors that critically influence large-scale properties. With a model, one can tune each factor individually to learn its affect on the assembly process. In this way, models can be used as a predictive guide to design new experiments.
 However, whether at atomistic or coarse-grained resolution, models involve important simplifications or other inaccuracies in their representation of physical systems. Thus, comparison of model predictions to experiments is essential to identify and then refine important model limitations. Iterative prediction, comparison, and model refinement can identify the key  factors that govern assembly mechanisms. % eads to a greater understanding of which factors assembly mechanisms, but also which  factors confer robustness or sensitivity.

The large ranges of length and time scales (\AA--$\mu$m, ps--minute) that are relevant to most capsid assembly reactions hinder simulating the process with atomic resolution, although Freddolino et al. \cite{Freddolino2006} performed an all-atom simulation of the intact STMV virus.  Recently, approaches to systematically coarse-grain from atomistic simulations have been applied to interrogate the stability of intact viruses \cite{Arkhipov2006,Joshi2011a,Joshi2011} or to estimate subunit positions and orientations from cryo electron microscopy images of the immature HIV capsid \cite{Ayton2010}. All-atom molecular dynamics has been applied to specific elements of the assembly reaction \cite{Yu2012}. As we describe below, most efforts to model capsid assembly to date have considered simplified models which retain those aspects of the physics which are hypothesized to be essential; with the validity of the hypothesis to be determined by comparison of model predictions with experiments.

\section{ Thermodynamics of capsid assembly }
\label{sec:thermoSection}
We will begin our discussion of viral assembly by analyzing the formation process of an empty capsid. While this process is most relevant to viruses that first form empty capsids during assembly,   ssRNA capsid proteins have also been examined with \emph{in vitro} experiments in which the ionic strength and $p$H were adjusted to enable assembly of empty capsids.

\subsection{Driving forces}
\label{sec:drivingForces}
For assembly to proceed spontaneously, states with capsids must be lower in free energy than a state with only free subunits. The assembly of disordered subunits (and RNA or other components if applicable) into  an ordered capsid structure reduces their translational and rotational entropy, and thus must be driven by favorable interactions among subunits and any other components that overcome this penalty. We begin here with the protein-protein interactions; the subunit-RNA interactions that promote ssRNA capsids to assemble around their genome are discussed in section \ref{sec:cargo} and also reviewed in great detail by Siber, Bozic, and Podgornik \cite{Siber2012}.
Capsid proteins are typically highly charged and possess binding interfaces that  bury large hydrophobic areas. Thus, as with most protein-protein interactions \cite{Alberts2010}, capsid assembly results from a combination of hydrophobic, electrostatic, van der Waals, and hydrogen bonding interactions.  Covalent interactions typically do not play a role in assembly, although they appear during subsequent maturation steps for a number of viruses (e.g. the bacteriophage HK97 \cite{Wikoff2000}).

Importantly, all of these interactions are short-ranged under assembly conditions.
Van der Waals interactions and hydrogen bonds operate on length scales of a few angstroms. Electrostatic interactions are screened on the scale of the Debye length, $\lambda_\text{D} \approx 0.3/C_\text{salt}^{1/2}$, with $\lambda_\text{D}$ measured in nanometers and  the salt concentration $C_\text{salt}$ measured in molar units. At physiological ionic strength, $C_\text{salt}=0.15$ M,  the screening length is $\lambda_\text{D}\approx 1$ nm; \emph{in vitro} experiments typically occur within the range $[0.05,1]$ M. The hydrophobic interactions are similarly characterized by a length scale of approximately a $0.5-1$ nm \cite{Chandler2005, Patel2012, Yu2012}.

In many cases the interaction is primarily driven by hydrophobic interactions, attenuated by electrostatic interactions with directional specificity imposed by van der Waals interactions and hydrogen bonding at \AA\ length scales.
The importance of hydrophobic interactions and the sometimes antagonistic contributions of direct electrostatic interactions have been shown by measuring the dimerization affinity of the C-terminal domain of the HIV capsid protein under an extensive series of mutations to the dimerization interface \cite{Lidon-Moya2005,Alamo2003, Alamo2005}. Furthermore, Ceres and Zlotnick \cite{Ceres2002} showed that the  thermodynamic stability of HBV capsids   increases with both temperature and ionic strength. The  increase in stability with temperature suggests that hydrophobic interactions are the dominant driving force \cite{Chandler2005}. The increase in stability with ionic strength, on the other hand, suggests that the salt screens repulsive electrostatic interactions which  oppose protein association. Several models based on this hypothesis reproduce the dependence of protein-protein interaction strength on ionic strength measured in the experiments \cite{Kegel2006,Phillips2009,Siber2012}. However, it is worth noting that the experiments were performed on capsid protein with the highly charged C-terminal domain truncated, and it is difficult to pinpoint on the crystal structure which charges are responsible for repulsive interactions. Ceres and Zlotnick \cite{Ceres2002} suggested that higher salt concentrations could enhance assembly by favoring a capsid protein conformation which is active for assembly.

\subsection{Law of mass action}
\label{sec:thermo}
We now consider the assembly thermodynamics for subunits endowed with the interactions just described.
We begin by considering the equilibrium for a system of identical protein subunits assembling to form empty $T$=1 capsids. To make the calculation analytically tractable, we assume here that there is one dominant intermediate species for each number of subunits $n$; extending this assumption is conceptually straightforward. The word subunit refers to the basic assembly unit defined in section \ref{sec:introExperiments}.

\begin{figure}
\begin{center}
\epsfxsize=0.5\textwidth\epsfbox{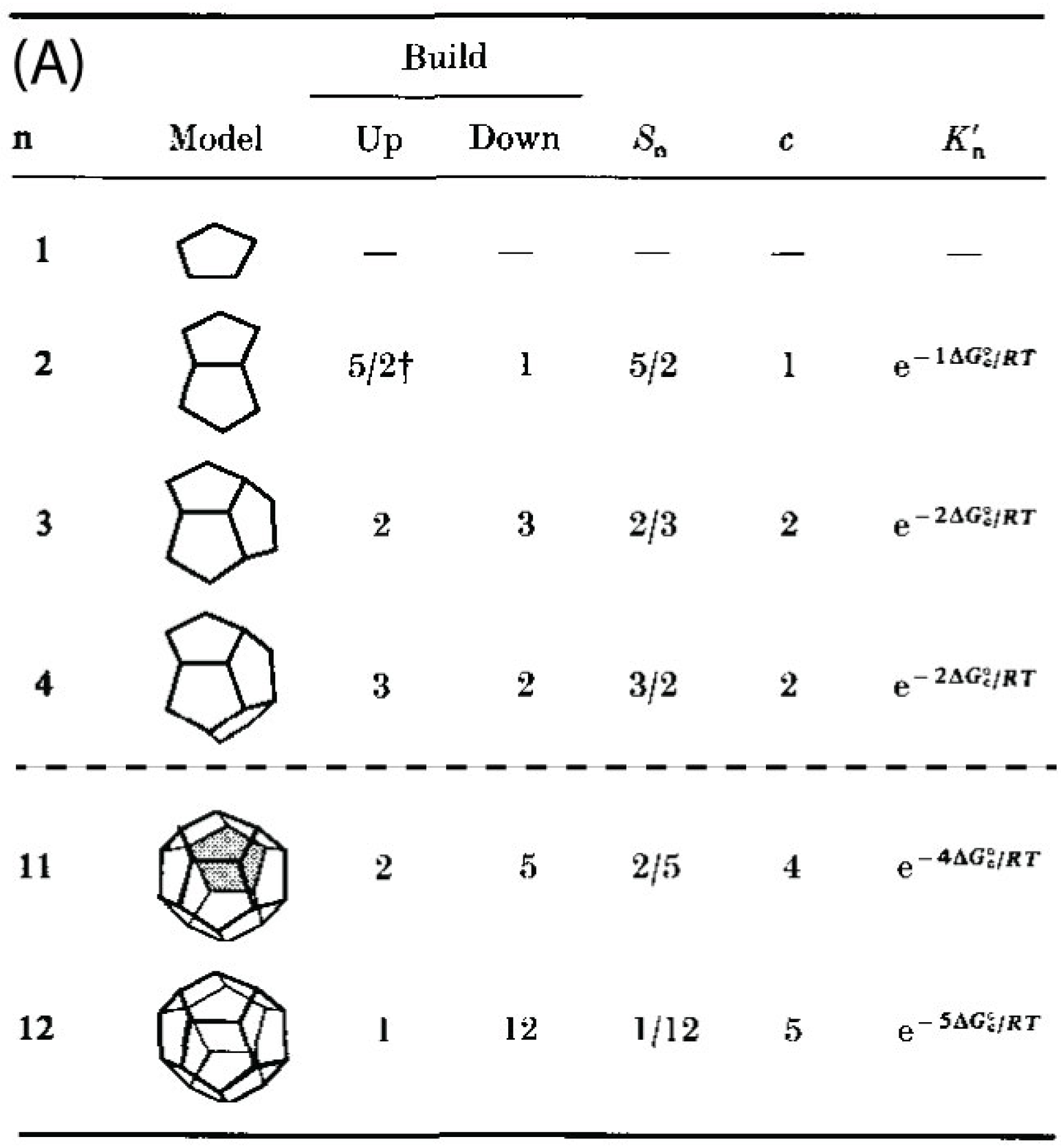}
\epsfxsize=0.5\textwidth\epsfbox{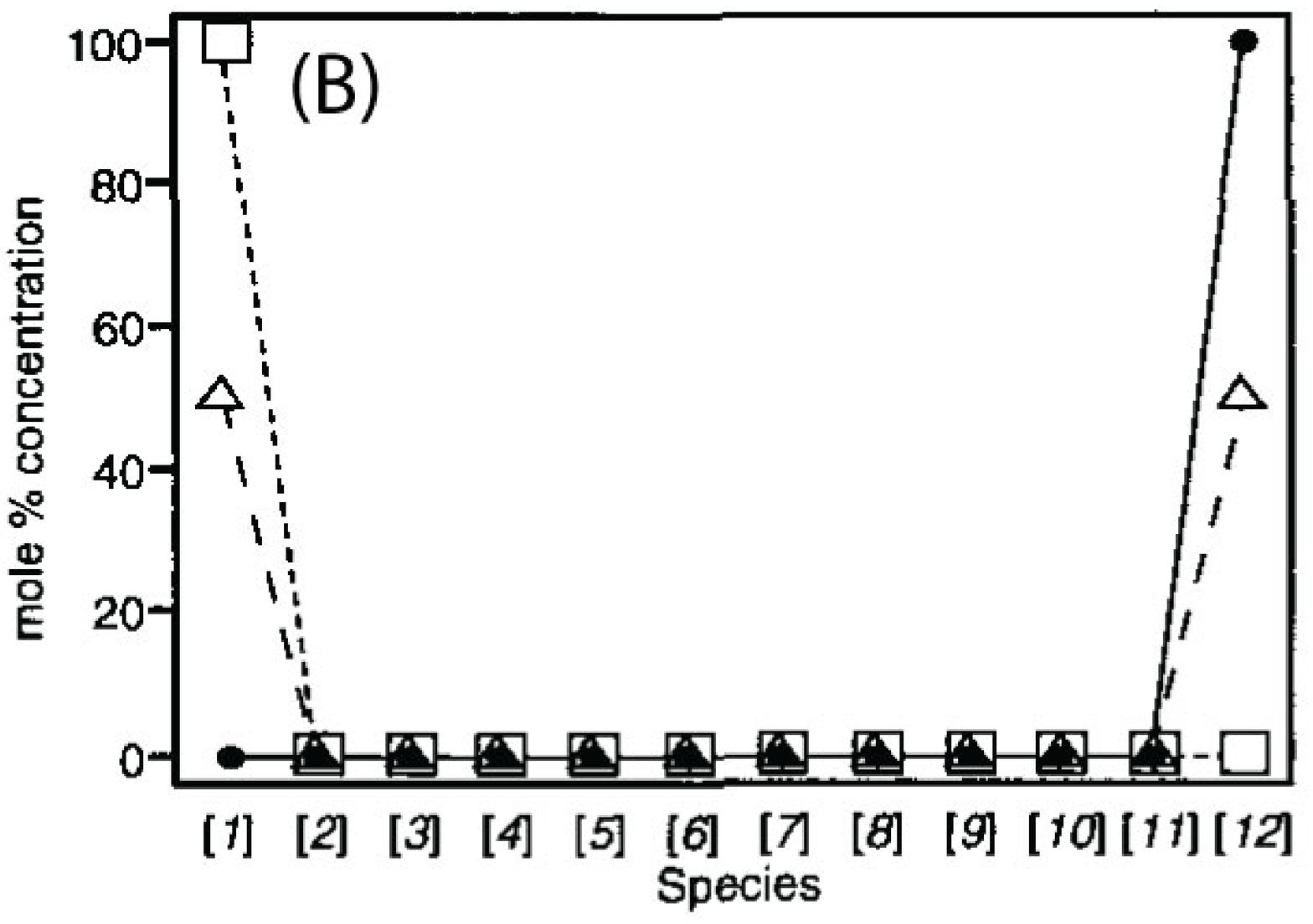}
\caption{{\bf (A)} The assembly model for a dodecahedral capsid and the statistical weights associated with symmetries for the intermediates. The columns list respectively the number of intermediates, the lowest energy configuration, the degeneracy for adding an additional subunit ($s_n$ in Eq.\ref{eq:ktEmpty} below), the degeneracy for losing a subunit ($\hat{s}_n$ in Eq.\ref{eq:ktEmpty}), the net degeneracy ($\sdegen_n$ in Eq. \ref{eq:Gi}), the number of contacts gained by adding a subunit($\ncontact_j$ in Eq. \ref{eq:Gi}), and the corresponding equilibrium constant. Only the first four and last two intermediates are shown; the full set are given in Ref. \cite{Zlotnick1994}. {\bf (B)} The mole fractions of each intermediate calculated using Eq. \ref{eq:rhon} and the statistical factors in (A) are shown for total subunit concentrations $\rhoTot$ of $0.44 \mu$M ($\Box$), $0.88 \mu$M ($\triangle$), and $1.8 \mu$M ($\CIRCLE$). Figures reprinted from J. Mol. Biol., {\bf 241}, Zlotnick, \emph{To Build a Virus Capsid: An Equilibrium Model of the Self Assembly of Polyhedral Protein Complexes}, 59-67 Copyright(1994)  with permission from Elsevier.
 \label{fig:dodecModel}
}
\end{center}
\end{figure}

The total free energy $\Fec$ for a system of subunits, intermediates, and capsids in solution can be written as
\begin{equation}
\Fec/\kt=\sum_{n=1}^N \kt \rho_n \left[\log(\rho_n v_0)-1\right] + \rho_n \Gcap_n
\label{eq:Fmt}
\end{equation}
where $v_0$ is a standard state volume, $\rho_n$ is the density of intermediates with $n$ subunits, and $\Gcap_n$ denotes the interaction free energy of such an intermediate. A plausible model for the interaction free energy is
\begin{equation}
\Gcap_n(\gb)=\sum_{j=1}^n (\ncontact_j \gb) - T \sdegen_n
\label{eq:Gi}
\end{equation}
where $\ncontact_j$ is the number of new subunit-subunit contacts formed by the binding of subunit $j$ to the intermediate, $\gb$ is the free energy for such a contact, and $\sdegen$ accounts for degeneracy in the number of ways subunits can bind to or unbind from an intermediate (see the $s$ factors in Refs.~\cite{Zlotnick1994, Endres2002} and Fig. \ref{fig:dodecModel}). These terms are specifed by the geometry of the capsid. Here we have subsumed rotational binding entropy penalties into $\gb$ (see Ref. \cite{Hagan2006,Hagan2011,Erickson1981,Ben-Tal2000}) and, to reduce the number of parameters, assumed that the binding energy $\gb$ is the same for all contacts.  As discussed in section \ref{sec:drivingForces}, $\gb$ depends on temperature, ionic strength, and $p$H.  Eq. \ref{eq:Gi} can be readily extended to allow for interface-dependent contact energies and subunit conformational changes \cite{Elrad2008}.

To obtain the equilibrium concentration of intermediates we minimize $\Fec$ subject to the constraint that the total subunit concentration $\rhoTot$ is conserved:
\begin{equation}
\sum_{n=1}^N n \rho_n = \rhoTot
\label{eq:rhoTot}.
\end{equation}
This yields the well-known law of mass action (LMA) result for intermediate concentrations \cite{Safran1994,Bruinsma2003}
\begin{align}
\rho_n v_0&=\exp[-\beta(\Gcap_n - n \mu)] \nonumber \\
\mu &= k_{\text{B}}T\log(v_0 \rho_1)
\label{eq:rhon}
\end{align}
with $\mu$ the chemical potential of free subunits and $\beta=1/\kt$. Due to the constraint (Eq. \ref{eq:rhoTot}) Eq. \ref{eq:rhon} must be solved numerically. The result for a model dodecahedral capsid comprised of 12 pentagonal subunits is shown in Fig. \ref{fig:dodecModel} for several values of the binding energy $\gb$. Notice that in all cases the capsid protein is almost entirely sequestered either as free subunits or in complete capsids. This prediction, which is analogous to the result for spherical micelles with a preferred diameter \cite{Safran1994} is generic to any description of an assembling structure in which the interaction free energy $\Gcap_n$ is minimized by one intermediate size ($n=N$) and  the total subunit concentration is conserved. %Under those two conditions, equilibrium must result in the chemical potential. It can be understood as follows. Association of the first pair of subunits makes only a single contact  to yield a favorable binding free energy $\gb$ but an entropy penalty of $\mu$. Because of the constraint (Eq. \ref{eq:rhoTot}) and the fact that larger intermediates have more contacts per subunit, at equilibrium ....  $\gb<\mu$.

\begin{figure}
\begin{center}
\epsfxsize=0.6\textwidth\epsfbox{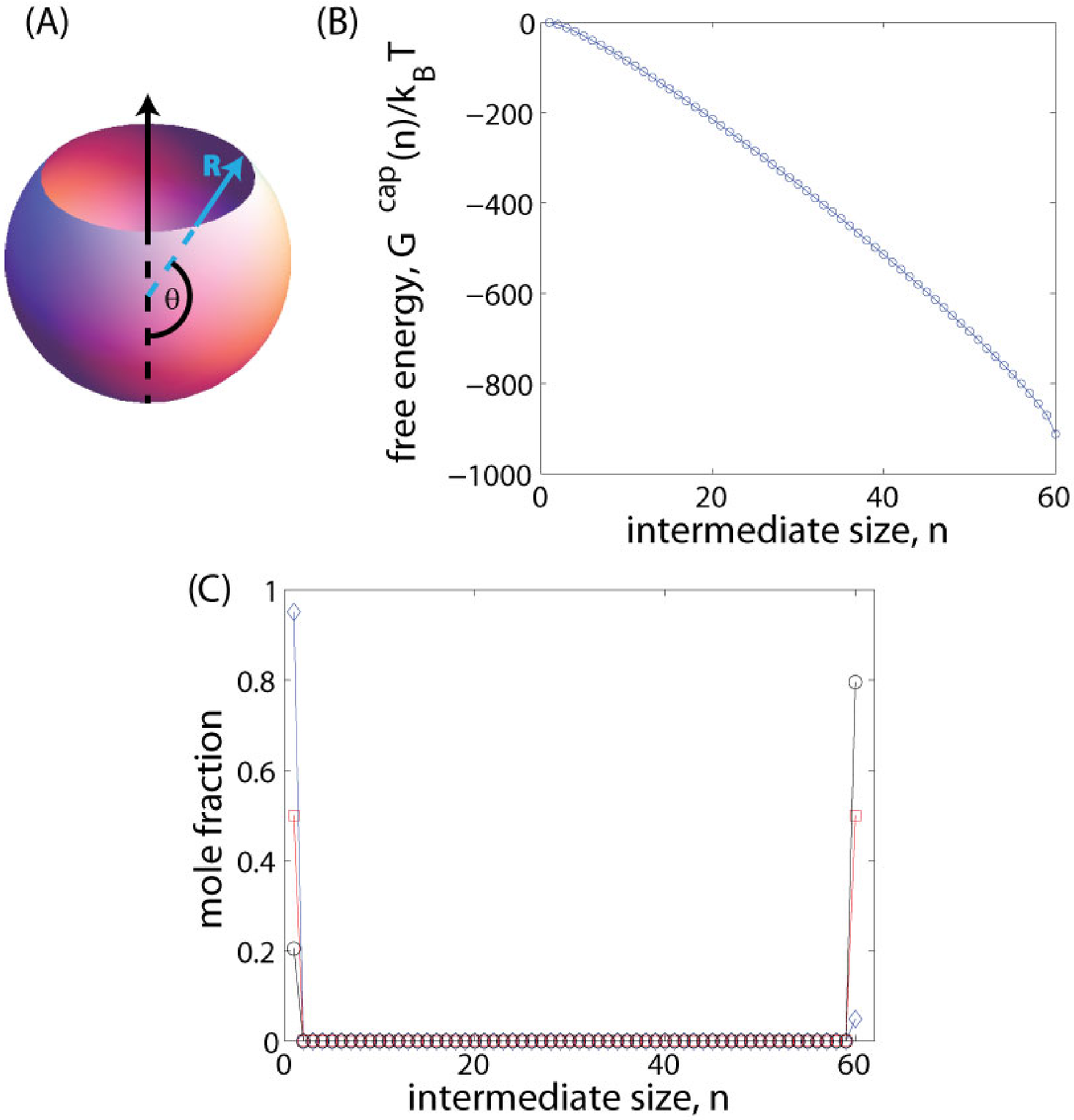}
\caption{ {\bf (A)} Depiction of the continuum model description of partial capsid intermediates considered by Zandi et al \cite{Zandi2006}.  $R$ is the radius of the capsid and the angle $\theta$ characterizes the extent of completion of the capsid. {\bf (B)} Interaction free energy $G(n)$ as a function of intermediate size $n$ obtained from Eq.\ref{eq:classNuc}. {\bf (C)} Predicted mole fractions using Eq.~\ref{eq:classNuc} and Eq.~\ref{eq:rhoTot} for $\rhoTot = \rhoStar$ ($\diamond$), $\rhoTot=2\rhoStar$ ($\square$), and $\rhoTot=5\rhoStar$ ($\circ$). $\gc = -15 \kt$ for (B) and (C).}
 \label{fig:zandiModel}
\end{center}
\end{figure}

To emphasize the generic nature of the prediction that intermediate concentrations are negligible at equilibrium, we also consider a  continuum model of an assembling shell presented by  Zandi and coworkers \cite{Zandi2006}. Each partial-capsid intermediate is described as a sphere, with a missing spherical cap. The unfavorable free energy due to unsatisfied subunit-subunit interactions at the perimeter of the cap is represented by a line tension $\sigma$, so the interaction free energy for a partial capsid with $n$ subunits is
\begin{align}
G^\text{cap}(n) = n \gc + \sigma l(n) - b % [i (N-i)]^{1/2}
\label{eq:classNuc}
\end{align}
with the perimeter of the missing spherical cap for a  given by
\begin{align}
l(n) = 2\pi R \sin \theta(n) = v_0^{1/3} 2 [\pi n(N-n)/N]^{1/2}
\label{eq:li}
\end{align}
with $v_0^{1/3}$ the size of one subunit and
$\gc$ the binding free energy per subunit (not per contact) in a complete capsid. Finally, we have included $b=\gc + 2 \sigma l(1)$ to ensure that free subunits have no interaction energy, since the continuum model breaks down for small intermediates. We set the line tension to $\sigma = -\gc/2$, which indicates that, on average, a subunit adding to the perimeter of the capsid satisfies half of its contacts. The resulting profile for $G(n)$ is shown in Fig.~\ref{fig:zandiModel}b, with the intermediate concentrations for several values of $\rhoTot/\rhoStar$ shown in Fig.~\ref{fig:zandiModel}c.  In all cases, the intermediate concentrations are negligible.
%We find that Eq.~\ref{eq:classNuc} with $\sigma=-\gc/2$ is a reasonable description during the growth phase in Brownian dynamics simulations, but fewer contacts are made by subunits associating to form the first polygon. Results shown in Fig.\ref{fig:zandiModel}.

\begin{figure}
\begin{center}
\epsfxsize=0.4\textwidth\epsfbox{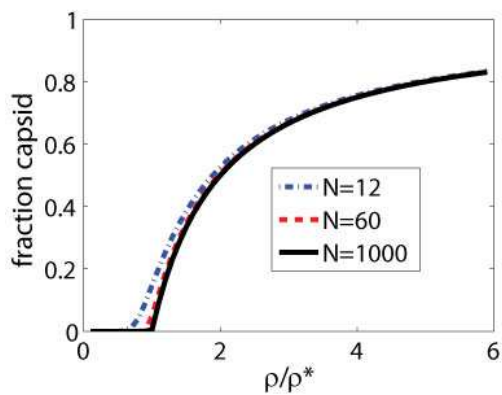}
\caption{ Fraction capsid $\fc$ as a function of subunit oversaturation $\rho/\rhoStar$ predicted by Eq. \ref{eq:fc} for the number of subunits in a complete capsid $N=12$, $60$, and $1000$.   }
 \label{fig:fVsPhi}
 \end{center}
\end{figure}

{\bf Two-state approximation.} Based on the observation that intermediate concentrations are negligible at equilibrium, the equations for capsid assembly thermodynamics can be simplified considerably by neglecting all intermediates except free subunits or complete capsids, so that
\begin{equation}
\rhoTot = \rho_1 + N \rho_N
\label{eq:rhoTottwo}.
\end{equation}
Defining the fraction of subunits in capsids as $\fc = N \rho_N/\rhoTot$, combining Eqs.~\ref{eq:rhoTottwo} and \ref{eq:rhon}, and rearranging, we obtain \cite{Schoot2007}
\begin{equation}
\frac{\fc}{1-\fc} = N (v_0 \phiTot)^{N-1} e^{-\beta \Gcap_N}
\label{eq:rhoTottwob}.
\end{equation}
In the limit $N\gg1$ this gives
\begin{align}
\frac{\fc^{1/N}}{1-\fc} & = \frac{\rhoTot}{\rhoStar} \nonumber \\
\rhoStar v_0 & =\exp\left(\beta \frac{\Gcap_N}{N-1}\right) N^{-1/(N-1)} \approx \exp\left(\beta \Gcap_N/N\right)
\label{eq:fc}
\end{align}
with $\rhoStar$ the pseudo-critical subunit concentration.
In the asymptotic limits  Eq. \ref{eq:fc} reduces to
\begin{align}
\fc & \approx \left(\frac{\rhoTot}{\rhoStar}\right)^N \ll 1 \qquad & \mbox{ for } \rhoTot\ll\rhoStar \nonumber \\
& \approx 1- \frac{\rhoStar}{\rhoTot} & \mbox{ for } \rhoTot\gg\rhoStar
\label{eq:fcAsympt}
\end{align}
 The solution to Eq. \ref{eq:fcAsympt} is shown in Fig. \ref{fig:fVsPhi} for several values of the capsid size $N$;
 note that the transition becomes sharper with increasing capsid size. Also notice that increasing the total subunit concentration $\rhoTot$ or the magnitude of the binding energy (i.e. decreasing $\rhoStar$) always increases the fraction of subunits in complete capsids $\fc$ at equilibrium. We will see however in section \ref{sec:dynamics} that this trend does not always apply at finite but experimentally relevant timescales due to kinetic effects.

{\bf Higher $T$ numbers.} If one or a few ground state capsid geometries are known (or pre-assumed), the thermodynamic calculation described above can be extended to describe  capsids with larger $T$ numbers in a straightforward manner. Recalling from section \ref{sec:anatomies} that icosahedral capsids comprise $T$ different subunit  conformations (or in some cases protein sequences), the capsid free energy $\Gcap_N$ must be extended to include conformation energies and contact free energies $\gb$ which depend on the subunit conformation or species \cite{Elrad2008}. Approaches to determine the lowest free energy configuration(s) for a shell are discussed in section \ref{sec:structuralStability}.

\subsubsection{Estimating binding energies from experiments}
\label{sec:estimateBindingEnergies}
\begin{figure}
\begin{center}
\epsfxsize=0.4\textwidth\epsfbox{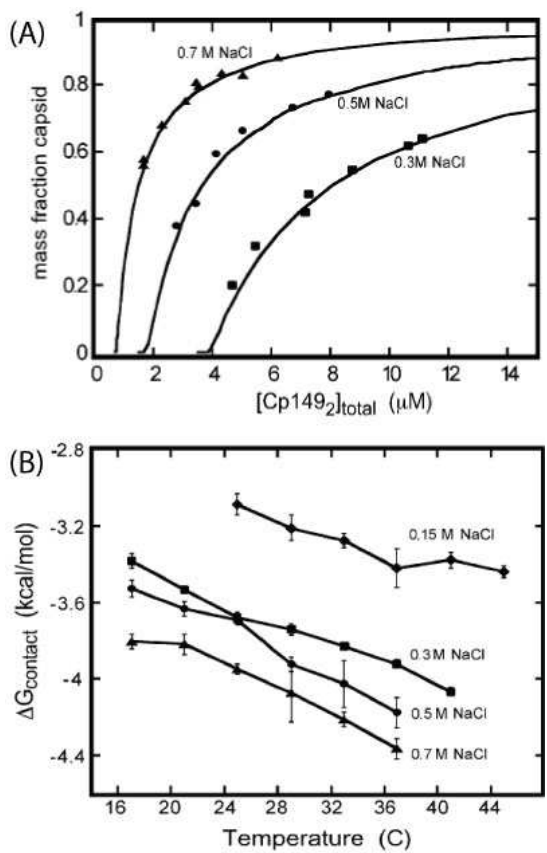}
\caption{  {\bf (A)} Fraction capsid measured for assembly of empty HBV capsids from capsid protein in which the RNA binding domain has been truncated, Cp149, using SEC as a function of total dimer subunit concentration [CP149]$_\text{total}$. Results are shown for indicated salt concentrations, and the lines are fits to the equilibrium model with $\Gcap_N=240 \gb -T \log(s_N)$ assuming four contacts per subunit and using the contact energy $\gb$ as a salt concentration dependent fit parameter, with the symmetry number of the complete $T$=4 capsid as $s_N=2^{119}/120$ \cite{Ceres2002}. {\bf (B)} Estimated values of $\gb$ as a function of temperature and ionic strength. Reprinted with permission from Ceres and Zlotnick, J. Mol. Biol., {\bf 41}, 11525-11531 (2002), \emph{Weak Protein-Protein Interactions Are Sufficient To Drive Assembly of Hepatitis B Virus Capsids}, Copyright (2002) American Chemical Society.}
 \label{fig:fExpVsPhi}
 \end{center}
\end{figure}

 Zlotnick and coworkers have shown that the assembly of HBV \cite{Ceres2002} can be captured by Eq. \ref{eq:fc} using $\gb$ as a fit parameter (see Fig.~\ref{fig:fExpVsPhi}). These fits yield an important observation that the subunit-subunit binding free energies are quite small, on the order of $\gb=4$ kcal/mol ($6.7\kt$) for productive assembly reactions. The observation that capsid assembly is driven by weak interactions of this magnitude appears to be a general rule for capsid assembly \cite{Zlotnick2003}, for reasons discussed in section \ref{sec:dynamics}.

  The conclusion that most of the interactions driving capsid assembly are weak appears to be broadly valid. However, it is important to note that  Eq. \ref{eq:fc} is an equilibrium expression, and thus strictly applies  only on times exceeding any relevant reaction timescale.   We can immediately see that this condition is beyond the reach of many experiments by estimating the timescale for a single subunit to leave an assembled capsid. Consider a typical subunit-subunit association rate constant of $f=10^5 / \text{M}\cdot s$ (\cite{Zlotnick1999,Endres2002,Johnson2005}, see section \ref{sec:rateEquations}), and a typical binding free energy of $\gb=6.7 \kt$. Since the dimer subunits of HBV are tetravalent, the first subunit must break four contacts to dissociate, with a timescale of about  $t_\text{dissociate} \sim f \exp(4\gb/\kt)=50$ days. Similarly, we show in section \ref{sec:assemblyTimeScales} that the approach of assembly toward equilibrium must lead to ever increasing nucleation barriers. Based on dynamical assembly simulations, our group has estimated that the values of $\gb$ could be underestimated by about $\kt$ even for measurements taken at 24 hours due to this effect.

The actual timescales for subunit dissociation from  complete capsids can be estimated from experiments that labeled subunits to monitor exchange with complete P22 capsids \cite{Parent2007} as well as $T$=3 and $T$=4 HBV capsids \cite{Uetrecht2010a}. Subunit exchange on a period of days to months was indeed demonstrated for the P22 capsids  and a fraction of the subunits in the $T$=3 HBV capsids. However, no subunit exchange was observed for $T$=4 HBV capsids, even when temperature was decreased to $4^\circ$ C (recall that HBV is less stable at lower temperature). Similarly, Singh and Zlotnick \cite{Singh2003} measured substantial hysteresis for the dissociation of HBV capsids under denaturant.  These observations raise the possibility that there are some steps which are irreversible (at least on measurement timescales) in the assembly process. Irreversible steps late in assembly or during a post-assembly maturation process make sense from the perspective of virus replication, as they would extend the period of time over which the virus can remain complete in infinite dilution and unfavorable environments. Of course, there must be a mechanism to release the genome once  the virus has infected a host.

The existence of irreversible steps cannot be directly revealed by assembly data alone. It has been shown that, even if there are assembly steps which are irreversible (on relevant timescales) late in the assembly cascade, as long as most steps are reversible the assembly data can be fit to Eq.\ref{eq:fc} with an apparent value of $\gb$ reflecting the free energy of the reversible steps \cite{Zlotnick2007,Morozov2009,Hagan2010,Roos2010}. Similarly, comparison of the dynamical equations described in section \ref{sec:rateEquations} to kinetics data could only reveal the presence of irreversible steps on timescales exceeding the equilibration time associated with the reversible steps (e.g. $\gtrsim50$ days).

\section{Modeling Self Assembly Dynamics and Kinetics of Empty Capsids}
\label{sec:dynamics}
The experimental measurements of capsid assembly kinetics described in section \ref{sec:introExperiments} provide important constraints on models of capsid assembly kinetics. At the same time, they present an important opportunity for modeling; because only some intermediates can typically be characterized, models are essential to understand detailed assembly pathways. In this section we describe different  modeling approaches which have been used to predict or understand the assembly kinetics.

\subsection{Timescales for capsid assembly}
\label{sec:simpleModel}
We begin our description of capsid assembly kinetics by defining the potential rate limiting steps and presenting scaling estimates for their timescales. While our estimates are based on simplified models, we will see in the subsequent sections that many of the predictions remain applicable when additional details are accounted for.

\label{sec:nucleationElongation}
It was noted by Prevelige \cite{Prevelige1993} that assembly kinetics for icosahedral capsids  can be described in terms nucleation and elongation (or growth) timescales, closely analogous to crystallization. Nucleation  refers to formation of a `critical nucleus', or a structure  which has a greater than 50\% probability of growing to a complete capsid before disassembling. Elongation then refers to the timescale required for a critical nucleus to assemble into a complete capsid. In contrast to crystallization, there can be a well-defined elongation timescale since capsids terminate at a particular size.

\begin{figure} [bt]
\begin{center}
\epsfig{file=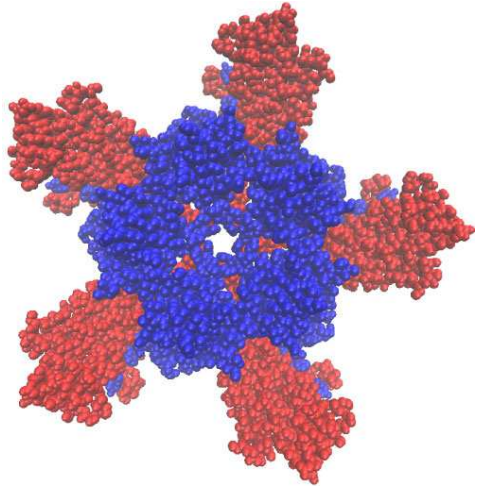,width=.35\textwidth}
\caption{Image of the CCMV pentamer of dimers that experiments \cite{Zlotnick2000} indicate is the critical nucleus. Atoms are shown in van der Waals representation and colored according to their quasi-equivalent conformation, with A monomers in blue and B monomers in red. The coordinates were obtained from the CCMV crystal structure, PDBID 1CWP \cite{Speir1995} using the Viper oligomer generator \cite{Reddy2001} and the image was generated with VMD \cite{Humphrey1996}. }
\label{fig:ccmvNucleus}
\end{center}
\end{figure}

%\subsubsection{Nucleation}
%\label{sec:nucleation}
{\bf Nucleation.} For any type of spheroidal shell, including an icosahedral capsid, the first subunits to associate create fewer interparticle contacts than those associating with larger partial capsids (see Figs.~\ref{fig:dodecModel}A and \ref{fig:zandiModel}B). Under conditions which lead to productive assembly the subunit-subunit binding free energy ($\gb$) is weak (see Fig.~\ref{fig:fExpVsPhi}). Thus the favorable free energy of these contacts is insufficient to compensate for the mixing and rotational entropy penalties incurred upon association, and the small intermediates are unstable. However, at subunit concentrations above the pseudocritical concentration $\rhoStar$ there must exist a critical size above which there are sufficient interactions such that further assembly is more probable than dissociation. In fact, the number of interactions depends on the partial capsid geometry, and thus there is an ensemble of critical nuclei with different sizes.

It is often assumed that the dominant assembly pathways pass through one or a few critical nuclei with the smallest sizes and thus the assembly probability can be approximated by a single valued function of partial capsid size $n$ (i.e. $n$ is a good reaction coordinate \cite{Dellago2002}).
Then, the critical nucleus corresponds to a maximum in the grand free energy, defined as $\Omega_n=G_n-\mu n$, with $G_n$ the Gibbs free energy for an intermediate with $n$ subunitsn and $\mu$ the chemical potential.
For the thermodynamic models of partial capsids presented in section \ref{sec:thermo} the grand free energy is given by
 \begin{equation}
 \Omega_n = \Gcap_n - \kt n \log(\rho_1 v_0)
 \label{eq:omegaNucleation}.
 \end{equation}
   with $\Gcap_n$ the interaction energy for a partial capsid intermediate with $n$ subunits and $\rho_1$ the free subunit concentration.

\begin{figure}[bt]
\begin{center}
\epsfig{file=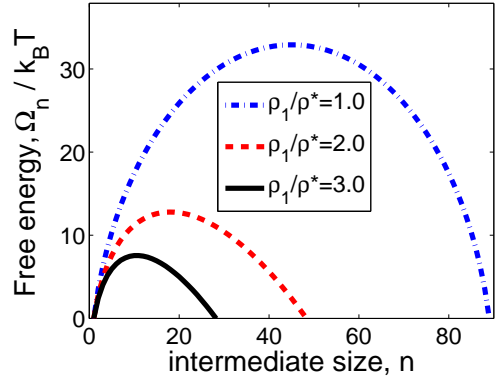,width=.4\textwidth}
\caption{ The grand free energy as a function of intermediate size for different free subunit supersaturation values ($\rho_1/\rho^*$) as calculated by the continuum model for a capsid with 90 subunits. These curves  would correspond to the free energy profiles at increasing times for a reaction which begins with $\rho_1=3\rho^*$ and proceeds toward equilibrium with $\rho_1=\rho^*$. }
\end{center}
\label{fig:nucleationBarriers}
\end{figure}

The effect of the shell geometry on the critical nucleus size can be understood elegantly from the continuum model of Zandi et al.\cite{Zandi2006}  in which partial capsid intermediates are described as spheres missing hemispherical caps with the partial capsid interaction free energy $\Gcap$ given by Eq.~\ref{eq:classNuc}. The critical nucleus is then calculated as the maximum in $\Omega(n)$ (Eq. \ref{eq:omegaNucleation}) to give \cite{Zandi2006}
\begin{align}
\nnuc = 0.5 N \left(1-\frac{\Gamma}{(\Gamma^2+1)^{1/2}}\right)
\label{eq:critNucCNT}
\end{align}
with $\Gamma = [\gc - \ln (\rho_1 v_0)]/\sigma$.
Notably, the critical nucleus size depends on the binding energy and subunit concentration, and  decreases with  increasing  supersaturation of free subunits ($\rho_1/\rhoStar$). Plots of $\Omega(n)$ are shown for several values of supersaturation in Fig.~\ref{fig:nucleationBarriers}.

The free energy forms for models which account for the icosahedral geometry of capsid structures are similar to the continuum model just described, except that the critical nucleus tends to correspond to a small polygon, which is a local minimum in the free energy since it corresponds to a local maximum in the number of subunit-subunit contacts (see Fig.~\ref{fig:dodecModel}). Although the assumption that there is one dominant intermediate per partial capsid size is an oversimplification, simulations \cite{Hagan2011,Rapaport2010} and theory \cite{Moisant2010,Endres2005} indicate that under many conditions assembly pathways predominantly pass through only a few nucleus structures which correspond to completion of small polygons. Measured critical nucleus sizes under simulated conditions have ranged from 3-10 subunits \cite{Elrad2008, Kivenson2010, Elrad2010,Rapaport2010}.

 Experimentally, nucleation has also been shown to correspond to completion of polygons, such as  the pentamer of dimers for CCMV \cite{Zlotnick2000} shown in Fig. \ref{fig:ccmvNucleus} or a trimer of dimers for turnip crinkle virus \cite{Sorger1986}.  However, it is likely that intertwining of flexible terminal arms and other subunit conformation changes can provide additional stabilization upon polygon formation. In the case of MS2, mass spectrometry \cite{Stockley2007} identified two polygonal intermediates, which modeling \cite{Morton2010} suggested were each critical nuclei for a different assembly pathway, with the prevalence dictated by binding to RNA.
  Most computational simulations of icosahedral capsids to this point have not incorporated allostery. Including stabilization due to polygon- or RNA-associated allostery could enable a particular structure to remain as the predominant critical nucleus over a wider range of interaction strengths and subunit concentrations than is predicted by more basic models.

%\subsubsection{Elongation}
%\label{sec:elongation}
{\bf Elongation.}
The association of subunits after nucleation has been described as elongation or growth. In contrast to the transient intermediates found below the critical nucleus size(s), intermediates in the growth phase are stable. Simulations indicate that association usually proceeds by the sequential addition of one or a few subunits at a time, although binding of larger oligomers can be significant at high concentrations or for some subunit interaction geometries \cite{Zhang2006,Hagan2006,Sweeney2008}. Association of large oligomers can also misdirect the assembly process \cite{Whitelam2009} (section \ref{sec:particleBased}).

\subsubsection{Scaling estimates for assembly timescales}
\label{sec:assemblyTimeScales}
To facilitate the presentation of how the timescales of nucleation and growth depend on system parameters, we first consider a highly simplified assembly reaction. It was shown that the conclusions from this simplified reaction remained valid when  more sophisticated models were considered\cite{Hagan2010}.

We consider a system of capsid protein subunits with total concentration $\rhoTot$ that start assembling at the time $t=0$. Our reaction is given by:
\begin{equation}
%1\xrightleftharpoons[b_\text{nuc}]{f c_1} 2 \xrightleftharpoons[b_\text{nuc}]{f c_1} \cdots \xrightleftharpoons[b_\text{nuc}]{f c_1} n_\text{nuc} \xrightleftharpoons[b_\text{elong}]{f c_1}\cdots \xrightleftharpoons[b_\text{elong}]{f c_1} N-1 \xrightleftharpoons[b_\text{N}]{f c_1}N
1\xrightleftharpoons[b_\text{nuc}]{f \rho_1} 2 \xrightleftharpoons[b_\text{nuc}]{f \rho_1} \cdots \xrightleftharpoons[b_\text{nuc}]{f \rho_1} n_\text{nuc} \xrightleftharpoons[b_\text{elong}]{f \rho_1}\cdots  \xrightleftharpoons[b_\text{elong}]{f \rho_1}N
\label{eq:zloteq}
\end{equation}
where $N$ is the number of subunits in a capsid, $\rho_1$ is concentration of unassembled subunits, $b_i$ is the dissociation rate constant (with $i=\{\text{nuc,elong}\}$), which is related to the forward rate constant by the equilibrium constant, $v_0 b_i=f \exp\left(\beta g_i\right)$, with $g_i$ the change in interaction free energy upon subunit association to a partial capsid and $v_0$ the standard state volume.  The nucleation and elongation phases are distinguished by the fact that association in the nucleation phase is not free energetically favorable, $\rho_1 \exp(-\beta \gnuc)<1$, while association in the elongation phase is favorable, $\rho_1 \exp(-\beta \gelong)>1$. Similar results can be obtained by assuming that the forward rate constant differs between the two phases \cite{Zlotnick1999}. For the moment, we assume that there is an average nucleus size $\nnuc$. %The  factors that determine $\nnuc$ are discussed in section \ref{sec:nucleationGrowth}.

We write the overall capsid assembly time $\tau$ as
 \begin{equation}
 \tau = \tau_\text{nuc} + \telong
  \label{eq:assemblyTime},
  \end{equation}
  with $\tau_\text{nuc}$ and $\telong$ the average times for nucleation and elongation, respectively. The timescale for the elongation phase can be calculated as the mean first passage time for a biased random walk with reflecting boundary conditions at $\nnuc$ and absorbing boundary conditions at $N$, with forward and reverse hopping rates given by $f \rho_1$ and $b_\text{elong}$, respectively. This gives  \cite{Bar-Haim1998}
\begin{equation}
\telong=\frac{\nelong}{f \rho_1 - b_\text{elong}} - \left(\frac{b_\text{elong}}{f \rho_1-b_\text{elong}}\right)^2\left(\frac{b_\text{elong}}{f \rho_1}\right)^{n_\text{elong}}
\label{eq:mfpt}
\end{equation}
with $\nelong=N-\nnuc$. In the limit of $f \rho_1 \gg b_\text{elong}$ Eq.~\ref{eq:mfpt} can be approximated to give $t_\text{elong}\approx \nelong/f \rho_1$, while for similar forward and reverse reaction rates, $f \rho_1 \approx b_\text{elong}$, it approaches the solution for an unbiased random walk $t_\text{elong} \approx \nelong^2/2 f \rho_1$.

Under conditions of constant free subunit concentration $\rho_1$, we can derive the average nucleation time with an equation analogous to Eq.~\ref{eq:mfpt} ~\cite{Hagan2008,Endres2002}
\begin{align}
\tnuc =&\frac{\nn}{f \rho_1 - b_\text{nuc}} - \left(\frac{b_\text{nuc}}{f \rho_1-b_\text{nuc}}\right)^2\left(\frac{b_\text{nuc}}{f \rho_1}\right)^{\nn} \nonumber \\
\approx & f^{-1}\exp\left(\Gcap_{\nn} /\kt \right) \rho_1^{-\nnuc} =
 1/f\rho_1\exp\left(-\Omega_{\nn} /\kt \right)
\label{eq:tnuc}
\end{align}
  This equation can be understood as follows. Because nucleation is rare on timescales of individual subunit binding events, pre-nuclei reach a quasi-equilibrium with concentration $\rho_n = \rho_1^n \exp[-\beta \Gcap_n]$ (see \eqref{eq:rhon}). The nucleation rate, $\tnuc^{-1}$ is then given by the rate of subunits adding to the largest pre-nucleus $\nn$, $\tnuc^{-1} = f \rho_1 \rho_{\nn}$ which gives the second line of \eqref{eq:tnuc}. A comparable expression is derived under a continuum limit in Ref. \cite{Zandi2006} in which the timescale for a subunit to associate with the critical nucleus is replaced by a critical nucleus survival timescale.

\begin{figure} [bt]
\begin{center}
\epsfig{file=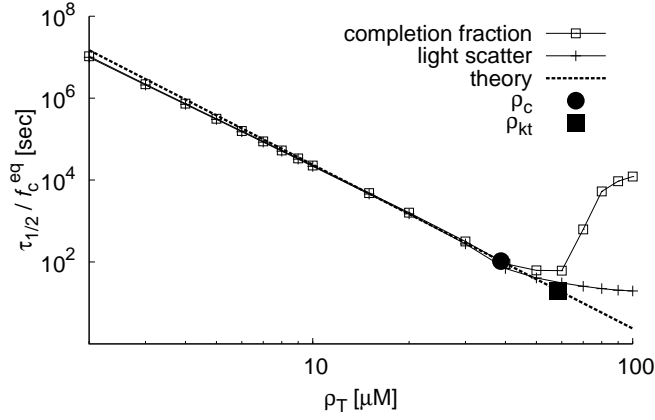,width=.5\textwidth}
\caption{ The scaling expression for the median assembly time $\halftime$ as a function of subunit concentration predicted by Eq. \ref{eq:halftime} is compared to full numerical solutions of the rate equations Eq. \ref{eq:zloteq} (see section \ref{sec:rateEquations}).  The numerical results are shown for completion fraction $\fc$ ($\Box$) and estimated light scattering ($+$), while the theoretical prediction  Eq. \ref{eq:halftime} is shown as a dashed line.  The estimate for the crossover concentration $\cc$ (Eq. \ref{eq:ckt}) above which the light scattering and completion fraction do not match is shown with a $\bullet$ symbol, and the concentration at which the monomer starvation kinetic trap increases overall assembly times $\ckt$ is shown as a $\blacksquare$ symbol. Parameter values are $\gnuc=7 \kt$ ($\approx 4$ kcal/mol) \cite{Ceres2002}, $\gelong=2\gnuc$, $g_N=4\gnuc$, capsid size $N=120$ corresponding to 120 dimer subunits in a Hepatitis B Virus capsid \cite{Ceres2002}, the critical nucleus size $\nnuc=5$, and the subunit association rate constant $f=10^5$ M$^{-1}$s$^{-1}$ \cite{Johnson2005}.	 Based on data from Ref. \cite{Hagan2010}.}
\label{fig:halfTimes}
\end{center}
\end{figure}

 However, because free subunits are depleted by assembly, the net nucleation rate never reaches this value and asymptotically approaches zero as the reaction approaches equilibrium. Instead, treating the system as a two-state reaction with $\nnuc$-th order kinetics yields an approximation for the median assembly time $\halftime$, the time at which the reaction is 50\% complete \cite{Hagan2010}
\begin{align}
\halftime \approx \frac{2^{\nn}-1}{\nn} \frac{\fceq}{N f}  \exp\left(G_{\nn}/\kt\right) \rho_0^{-\nn}
\label{eq:halftime}
\end{align}
with $\fceq$ as the equilibrium fraction of subunits in complete capsids, which can measured experimentally \cite{Ceres2002}. The factor of $N^{-1}$ in Eq.~\ref{eq:halftime} accounts for the fact that $N$ subunits are depleted by each assembled capsid. This prediction is compared to simulated assembly times in Fig.~\ref{fig:halfTimes}.

{\bf Kinetic trap.}  We found that the relationships between $\telong$ and Eq.~\ref{eq:halftime} and assembly times begin to fail at a crossover concentration $\cc$ for which the initial rate of subunit depletion by  nucleation ($N/\tnuc$) is equal to the elongation rate. For larger subunit concentrations, new nuclei form faster than existing ones complete assembly, and  free subunits are depleted before most capsids finish assembling.  The system then becomes kinetically trapped at a larger concentration $\ckt$ defined by the point at which the median assembly time $\halftime$ matches the elongation time.  These concentrations are related to binding free energies and other parameters by
\begin{align}
 \telong & \approx \tnuc/N &  \quad \mbox{for}\quad \rhoTot = \cc \nonumber \\
%& \approx  f^{-1}\exp\left(G_{\nn} /\kt \right) c_0^{-\nn}  \quad \mbox{for}\quad c_0=\cc \nonumber \\
\telong &\approx \halftime & \quad \mbox{for} \quad \rhoTot=\ckt
\label{eq:ckt}
\end{align}
with $\tnuc$ and $\halftime$ respectively given by Eq.~\ref{eq:tnuc} and  Eq.~\ref{eq:halftime}.

 A kinetic trap arising from depletion of free subunits (Eq.\eqref{eq:ckt}) was first noted by Zlotnick \cite{Zlotnick1994,Zlotnick1999,Endres2002} and was observed in experiments on CCMV \cite{Zlotnick2000} and HBV \cite{Zlotnick1999} (see the largest ionic strength in Fig. \ref{fig:lightScatter}A).  Morozov, Bruinsma, and Rudnick \cite{Morozov2009} elegantly  recast a  model similar to Eq. \eqref{eq:zloteq} in a continuum description, within which the time evolution of concentrations of capsid intermediates resembles a shock wave.  If the wave does not reach the size of a complete capsid before free subunits are depleted then the system is trapped.

While the continuum model correctly predicts the presence of the free subunit depletion trap, the computer simulations described in sections \ref{sec:rateEquations} and \ref{sec:particleBased} show that productive capsid assembly reactions do not resemble a shockwave. Because nucleation is a stochastic event, each capsid elongation process starts at a different time; i.e., they are out of phase.  For $\rhoTot<\ckt$ relatively few capsids are assembling at any given time,  and thus intermediate concentrations remain at low levels. The shockwave could only arise in the limit of $\tnuc \ll \telong$, in which case the system would  be severely trapped. This trap can be avoided though for reactions in which subunits assemble around RNA or nanoparticles (section \ref{sec:cargo}), provided that subunits are in excess. %It can be seen from Eq.\eqref{eq:telong} and computer simulations that the distribution of elongation times has a small standard deviation; if the elongation process  were in phase it would resemble a shockwave.

\subsubsection{Lag times}
\label{sec:lagTimes}
A distinctive feature of many capsid assembly kinetics measurements is an initial lag phase before detectable assembly occurs (e.g. Fig. \ref{fig:lightScatter}) whose duration decreases with increasing subunit concentration or subunit-subunit binding free energy.
Although Zlotnick and coworkers \cite{Zlotnick1994,Zlotnick1999,Endres2002} showed that partial capsid intermediates assemble during the lag phase, it has often been assumed that the duration of the lag phase corresponds to the time required for the concentration of critical nuclei to reach steady state, in analogy to models of actin nucleation. However, the simple model described above can be solved exactly in the limit of constant free subunit concentration \cite{Hagan2008,Hagan2010}, in which case the lag phase in $\fc$ is equal to the mean elongation time $\telong$ estimated in the previous section. Because the free subunit concentration is nearly constant during the lag phase under most conditions, this relationship holds even when the assumption of constant free subunit concentration is relaxed.

\begin{figure} [bt]
\begin{center}
\epsfig{file=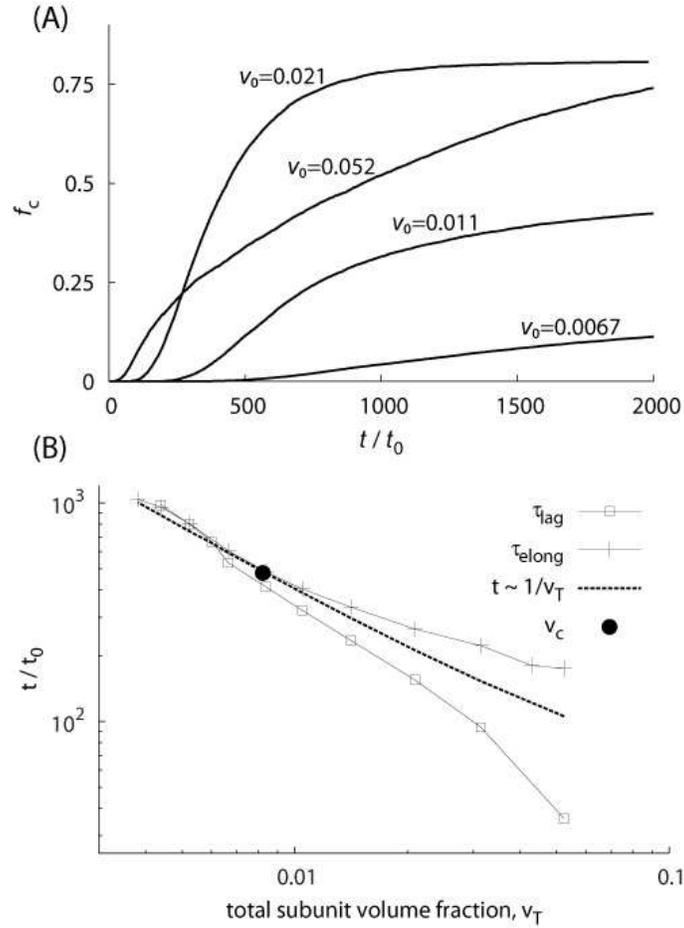,width=.5\textwidth}
\caption{ The lag time is related to the mean elongation time. {\bf (A)} Completion fractions $\fc$ measured from Brownian dynamics simulations of a particle based model (section \ref{sec:particleBased}) are shown as a function of time for indicated total subunit volume fractions ($v_\text{T}$). {\bf (B)} The duration of the lag phases from the simulations shown in (A) are compared to  mean capsid elongation times.   The crossover volume fraction $v_\text{c}$ estimated from Eq. \ref{eq:ckt} is shown as a $\bullet$ symbol. The plotted data is from Ref. \cite{Hagan2010}.}
\label{fig:lagTimes}
\end{center}
\end{figure}

To illustrate this relationship, mean elongation times and lagtimes calculated from Brownian dynamics simulations of a particle-based model (see section \ref{sec:particleBased} and Ref \cite{Hagan2010}) are shown in Fig. \ref{fig:lagTimes}. We see that the correspondence is excellent until the reaction approaches the crossover concentration $\cc$ (estimated from Eq. \ref{eq:ckt}).
This correspondence could be tested experimentally by comparing elongation times measured by single molecule experiments \cite{Jouvenet2008,Baumgaertel2012,Jouvenet2011} with lag times measured by bulk experiments.  %New experimental techniques such as the resistive pulse sensing apparatus \cite{Zhou2011,Harms2011} may facilitate experimental measurement of lag times.

%In both of these models the elongation time and hence the duration of the lag phase scale inversely with the free subunit concentration because the elongation is primarily a first-order reaction. Higher order elongation reactions are possible \cite{Schwartz} in which case the scaling with free subunit concentration should change but the relationship between lag time and elongation time should not. However, the above analysis does not consider protein subunit conformational transitions between active and inactive states \cite{Chen2008,Zlotnick}; if the timescale for these is sufficiently long in comparison to the elongation time it will also be reflected in the measured kinetics. This relationship has not yet been tested experimentally in part due to the difficulty associated with accurately measuring the duration of the lag phase. However, new experimental techniques such as the resistive pulse sensing apparatus \cite{Zhou2011,Harms2011} may facilitate experimental measurement of lag times.

\subsubsection{The slow approach to equilibrium}
 \label{sec:approachEquil}
 To this point in this section we have made the simplifying assumption that there is a fixed critical nucleus size. However, Eq. \ref{eq:critNucCNT} shows that in general the critical nucleus size is a function of the free subunit concentration. As subunits are converted into capsids by the reaction, the free subunit concentration ($\rho_1$) decreases and hence the nucleation barrier grows. In Fig. \ref{fig:nucleationBarriers} nucleation barriers calculated using Eqs. \ref{eq:classNuc}, \ref{eq:li}, and \ref{eq:omegaNucleation} for a capsid with $N=90$ subunits are shown at three time points (i.e. three free subunit concentrations $\rho_1$) for a reaction which begins with a supersaturated free subunit concentration $\rho_1(t=0)=3\rho^*$ with $\rho^*$ the pseudocritical subunit concentration (section \ref{sec:thermo}). As the reaction begins far out-of-equilibrium there is a relatively small critical nucleus size and correspondingly a small nucleation barrier. However, as the reaction approaches equilibrium $\rho_1=\rho^*$ the size increases to a half-formed capsid and the barrier increases to $30\kt$. Substitution of this free energy barrier into Eq.\ref{eq:tnuc} shows that the reaction timescale far exceeds the experimentally accessible timescales at this point. In other words the reaction only approaches equilibrium asymptotically. As noted in section \ref{sec:thermo}, this effect can lead to underestimating subunit-subunit binding energies when finite-time assembly data is fit to equilibrium expressions.

The observation that, {\bf at equilibrium}, the critical nucleus corresponds to a half capsid $\nnuc=N/2$ is rather generic. It results from the fact that equilibrium is reached when the free subunit concentration decreases to the point at which the chemical potential of a free subunit is equal to that of a subunit in a complete capsid, $\rho_1^\text{eq} = \exp\left(\beta \Gcap_N/N\right)$. Thus, including additional  complexities, such as subunit conformational changes or interface-dependent binding energies would not qualitatively change the result. We reiterate, though, that for the supersaturation conditions $\rho_1>\rho^*$ from which productive assembly begins, the critical nucleus size will generally be much smaller than its equilibrium value.

\subsection{Rate equation models for capsid assembly}
\label{sec:rateEquations}
Zlotnick and coworkers \cite{Zlotnick1994,Zlotnick1999,Endres2002} developed an approach to simulate empty capsid assembly via a system of rate equations that describe the time evolution of concentrations of empty capsid intermediates. The idea is analogous to the classic kinetic rate equations for cluster concentrations in a system undergoing crystallization proposed by Becker and D\"oring~\cite{Becker1935} and then derived from the microscopic dynamics of the lattice gas (or Ising) model by Binder and Stauffer~\cite{Binder1976}. In contrast to the models for crystallization, however, the capsids terminate at a finite size.
The initial works of Zlotnick and coworkers used the simplifications that there is one species of intermediate for each size $n$, and that only single subunits can bind or unbind in each step, which gives the following equations
\begin{eqnarray}
\frac{d \rho_1}{d t} &=& -f_1 s_1 \rho_1^2 + b_2 \hat{s}_2\rho_2 +\sum_{n=2}^{N}-f_n s_n \rho_n \rho_1 + b_n \hat{s}_n \rho_n \nonumber \\
\frac{d \rho_n}{d t}&=&f_{n-1} s_{n-1} \rho_1 \rho_{n-1} - f_n s_n \rho_1 \rho_n \qquad \qquad n=2\dots N \nonumber \\
& & -b_n \hat{s}_n \rho_n + b_{n+1} \hat{s}_{n+1} \rho_{n+1}
\label{eq:ktEmpty}
\end{eqnarray}
where $\rho_n$ is the concentration of intermediates with $n$ subunits, $f_n$ and $b_n$ are respectively association and dissociation rate constants for intermediate $n$, and $s_n$ and $\hat{s}_n$ respectively describe the degeneracy for binding and unbinding \cite{Endres2002}. While Eq. \ref{eq:ktEmpty} resembles a Master equation, notice that the factors of $\rho_1$ in the association reactions introduce nonlinearities which complicate its solution. Therefore we will refer to the equations in this section as `rate equations'.

The association and dissociation rate constants are related by detailed balance as $b_n=f_{n-1} \exp\left[(\Gcap_n-\Gcap_{n-1})/\kt\right]/v_0$, with $\Gcap_n$ the interaction free energy of a partial capsid with $n$ subunits and $v_0$ the standard state volume.  Specifying the assembly model requires defining the set intermediate geometries and their free energies (e.g. see section \ref{sec:thermo}, Eqs. \eqref{eq:Gi} or \eqref{eq:classNuc}) and the association rates ${f_n}$. %In many cases the system of equations is simplified by choosing average values for the degeneracy factors $s_n$ and $\hat{s}_n$, the association rate constants $f_n$ and the binding free energies over ranges of $n$, see e.g. section (previous section) or \cite{Endres2002}. %While this might seem like an oversimplification, it can actually improve the results. The assumption that

\begin{figure}
\begin{center}
\epsfxsize=0.50\textwidth\epsfbox{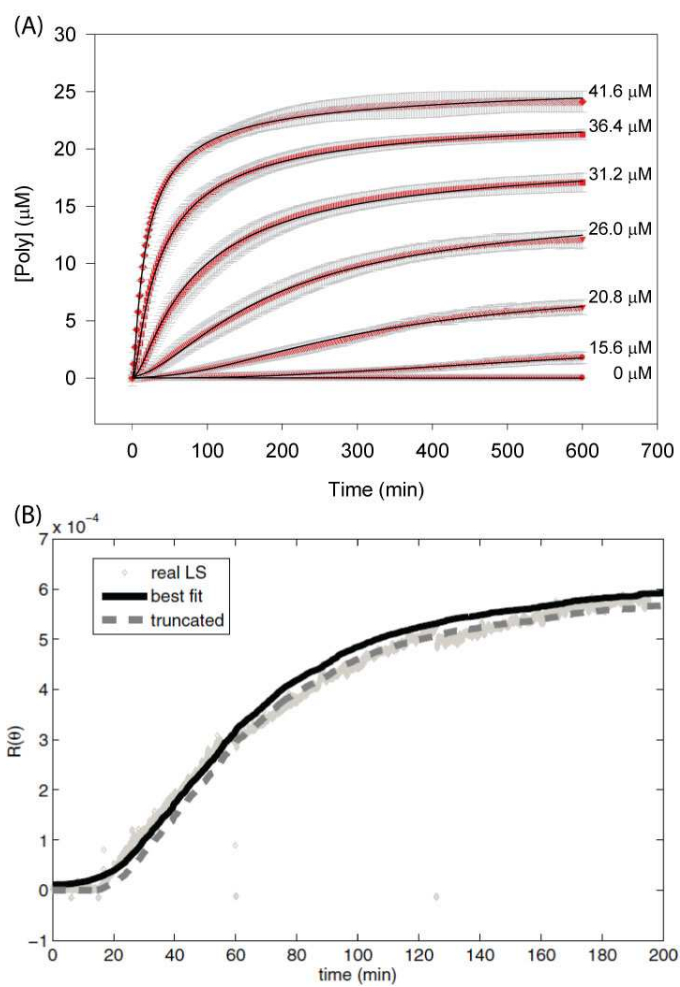}
\caption{{\bf (A)} \emph{In vitro} assembly of HIV CA protein into tubes monitored by absorbance (red diamonds, with thick grey lines indicating error bars) at indicated subunit concentrations compared to best fits using a rate equation model (black lines). {\bf (B)}    Light scattering for HPV LP1 assembly from Casini et al. \cite{Casini2004} (light grey diamonds) compared to a continuous time Monte Carlo trajectory using parameters optimized to the data (solid black line).  The dashed curve corresponds to a trajectory with parameter values reduced by $2.5\times10^5$ from their optimal values and negative values truncated to zero, to simulate a threshold level of signal to background scattering. (A) is reprinted with permission from Biochemistry, {\bf 51}, 4416-4428 (2012), \emph{A Trimer of Dimers Is the Basic Building Block for Human Immunodeficiency Virus-1 Capsid Assembly}, Tsiang, Niedziela-Majka, Hung, Jin, Hu, Yant, Samuel, Liu, Sakowicz, Copyright (2012) American Chemical Society. (B) is reprinted with permission from Phys. Biol., {\bf 7}, 045005 (2010), Kumar and Schwartz, \emph{A parameter estimation technique for stochastic self-assembly systems and its application to human papillomavirus self-assembly}, Copyright (2010) IOP Publishing. }
 \label{fig:KumarRateEquationsFitToData}
 \end{center}
\end{figure}

Despite the extreme simplifications leading to Eq. \ref{eq:ktEmpty}, rate equations of this form have shown good agreement with many features of experimental assembly kinetics data, including the assembly kinetics of HBV \cite{Zlotnick1999} (Fig.~\ref{fig:lightScatter}B), the assembly of CCMV into different polymorphs depending on subunit concentration \cite{Johnson2005}, the short time kinetics of BMV assembly \cite{Chen2008},  SV40 assembly \cite{Keef2006}, the impact of RNA on MS2 assembly \cite{Morton2010}, and the assembly of HIV capsid protein into tubes \cite{Tsiang2012} (Fig.~\ref{fig:KumarRateEquationsFitToData}A).

The assumption of only a single structure per intermediate size can be relaxed at the cost of increased computational complexity. For example Zlotnick and coworkers \cite{Endres2005,Moisant2010} have enumerated the space of all possible well-formed cluster configurations for two geometries, and catalogued the ensemble of pathways  surpassing a threshold value of probability \cite{Moisant2010}. In an alternative approach, Schwartz and coworkers \cite{Zhang2006,Sweeney2008} have used continuous time Monte Carlo (known as the Bortz-Kalos-Lebowitz \cite{Bortz1975} or Gillespie \cite{Gillespie1977} algorithm) to stochastically sample pathways consistent with kinetic rate equations. They have particularly considered the effects of binding between oligomers \cite{Zhang2006,Sweeney2008,Xie2012} and an optimization routine to fit parameters to light scattering data \cite{Kumar2010,Xie2012} (see Fig. \ref{fig:KumarRateEquationsFitToData}).

Several groups have also developed continuum-level descriptions of assembly dynamics which allow for analytical treatment \cite{Schoot2007,Morozov2009}. Van der Schoot and Zandi applied the classical theory of spinodal decomposition (model A dynamics) to examine the late-stage relaxation of assembly dynamics, while Ref. \cite{Morozov2009} is discussed in section \ref{sec:thermo}. However, the important role of nucleation in the kinetics has not yet been incorporated into these treatments.

{\bf Limitations and advantages of the rate equation approach.}
 The key advantage of state-based  over particle-based approaches discussed next is that the former do not track diffusive motions of individual subunits and thus can access  larger system-sizes and timescales. However, even extended state-based approaches require pre-definition of the accessible state space (i.e. the  structures of intermediates for each size $n$) and the transition rates between them. To date these methods have not been used to address the possibility of strained interactions between subunits which deviate from the ground state of the pairwise interaction potential. The possibility of strained interactions would greatly expand the set of possible cluster configurations, hindering predefinition of the state-space.  Secondly, the state-based approaches used to date assume a uniform  spatial distribution of free subunits (a mean-field  approximation), neglecting any particle-particle correlations or rebinding kinetics. However, there is no evidence from particle-based simulations or experiments that this approximation leads to significant error.

\subsection{Particle-based simulations of capsid assembly dynamics}
\label{sec:particleBased}
In this section we consider simulations of capsids or other polyhedral shells that explicitly track the positions and orientations of each subunit. Thus, once the model has been defined no further assumptions about pathways or the state space are required.
Capsid proteins typically have several hundred amino acids and assemble on time scales of seconds to hours.  Thus, simulating the spontaneous assembly of even the smallest icosahedral capsid with 60 proteins at atomic resolution  would entail an extreme computational demand \cite{Freddolino2006}. However, it has been shown that the capsid proteins of many viruses adopt folds with similar excluded-volume shapes, often represented as trapezoids \cite{Mannige2008}. Several groups have therefore developed models for subunits which, although highly simplified, retain the most important features. Namely, they have an excluded-volume geometry and orientation-dependent attractions designed such that the lowest energy structure is shell with icosahedral symmetry \cite{Schwartz1998,Hagan2006,Hicks2006,Nguyen2007,Wilber2007,Nguyen2008,Nguyen2009,Johnston2010,Wilber2009a,Wilber2009,Rapaport1999,Rapaport2004,Rapaport2008,Elrad2010,Hagan2011,Mahalik2012}

\begin{figure}
\begin{center}
\epsfxsize=0.5\textwidth\epsfbox{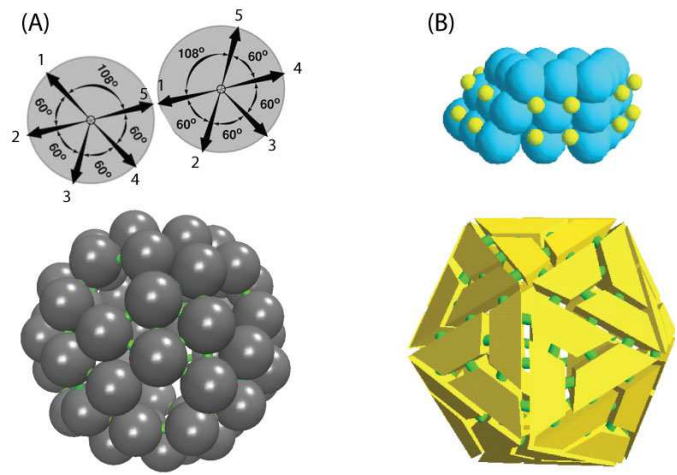}
\caption{Examples of two classes of models for icosahedral shells. {\bf (A)} A patchy-sphere model with the pentavalent subunit interaction geometry of a $T$=1 capsid (see Fig. \ref{fig:tNumber}C), but spherically symmetric excluded-volume \cite{Hagan2006}. In the top image, two interacting subunits are shown, with numbered arrows indicating the locations of the 5 distinct attractive patches. The lower image shows an assembled capsid, with patches colored green. {\bf (B)} An extended subunit representation of a $T$=1 capsid.  In the top image, the large cyan spheres experience repulsive excluded-volume interactions while small yellow spheres on complementary faces experience attractive interactions. The lower image shows a complete capsid, with subunits reduced in size for visibility and the locations of attractive patches indicated by green cylinders.  The images in (B) are reprinted with permission from Rapaport, Phys. Rev. E, {\bf 70}, 051905 (2004), \emph{Self-assembly of polyhedral shells: A molecular dynamics study}, Copyright (2004) by the American Physical Society.}
\label{fig:computationalModels}
\end{center}
\end{figure}

The coarse-grained particle-based simulation models can be roughly separated into three classes. We will use the term `patchy-sphere models' to refer to models in which the subunit has spherically symmetric excluded-volume and patches with short-ranged attractions arranged such that the lowest energy configuration corresponds to a particular target structure (see Fig. \ref{fig:computationalModels}A). Patchy-sphere models are quite general and have long been used to represent decorated colloids (e.g. \cite{Bianchi2006}) as well as proteins  (e.g. \cite{Liu2007}). The patch-patch potential can include angular and dihedral terms to control the overall directional specificity of the attraction \cite{Schwartz1998,Hagan2006,Wilber2009a} and patches with different interaction length scales to control preferred face angles of assembling polyhedrons \cite{Wales2005,Johnston2010}.

The second class of models, first developed by Rapaport \cite{Rapaport1999}, considers an extended subunit comprised of spherically symmetric `pseudoatoms' arranged to have short-ranged attractions and excluded-volume geometries that mimic features of protein geometries seen in capsid structures (Fig. \ref{fig:computationalModels}B). For example, several groups \cite{Rapaport2004,Rapaport2010,Nguyen2009} have considered models in which subunits have a trapezoidal shape which is roughly consistent with that of capsid proteins with the  beta-barrel architecture \cite{Rossmann1989} or models in which 20 triangular subunits (which could correspond to protein trimers)  form icosahedral shells \cite{Rapaport2004,Rapaport1999,Nguyen2007,Elrad2010,Hagan2011,Mahalik2012}.
 Extended subunits have also been used to model nanoparticles with a variety of shapes (e.g. \cite{Zhang2004a}).
 In a third class of models, subunits have polygonal interaction directions, but rather than tracking their diffusion, they are irreversibly placed onto growing capsids in energy-minimized configurations \cite{Hicks2006,Levandovsky2009}.

{\bf The early history of particle-based capsid assembly simulations.}
The first dynamical simulations of capsid assembly were performed by Schwartz and co-workers \cite{Schwartz1998}, who considered a patchy-sphere type model with complementary attractive interactions directed such that lowest energy configurations corresponded to 60-subunit $T$=1 closed shells. Their exploratory simulations using dissipative molecular dynamics identified the importance of annealing during assembly, as uncorrected assembly errors tended to lead to malformed structures. Rapaport considered models for icosahedral shells in which subunits have triangular \cite{Rapaport1999} or trapezoidal \cite{Rapaport2004} excluded-volume geometries.  The early simulations \cite{Rapaport2004} included dynamically unrealistic rules which limited the number of nucleation sites, but suggested that a simple interaction potential could direct assembly of well-formed capsids and that subunit association rates do not decrease dramatically as capsids near completion. %Examples of the patchy-sphere and extended subunit models with $T$=1 shells as the ground state are shown in Fig. \ref{fig:computationalModels}.      %Later simulations relaxed these rules \cite{Rapaport2007} and  included explicit solvent

The first statistical estimates of assembly into icosahedral shells were obtained by Hagan and Chandler \cite{Hagan2006}  using overdamped Brownian dynamics with several patchy-sphere models for $T$=1 shells. They constructed a `kinetic phase diagram' showing the dominant assembly products as a function of subunit concentration, subunit-subunit interaction strength, and the orientational specificity of subunit interactions (see below). They also found that assembled capsids were highly metastable in infinite dilution and disassembly showed significant hysteresis, as seen in experiments on HBV capsids \cite{Singh2003}. %Furthermore, association of oligomers was a frequent event for two of the interaction geometries studied.

Nguyen and co-workers \cite{Nguyen2007} used discontinuous molecular dynamics \cite{Alder1959,Rapaport1979,Rapaport1978,Bellemans1980,Smith1997} to simulate the assembly of models in which subunits had short-range attractive interactions and excluded-volume geometry shapes of triangles or kites. In contrast to other models in which subunits are rigid bodies, the pseudo-atoms comprising each subunit were connected by infinite hardwall potentials and thus the subunits had some internal degrees of freedom. They predicted a phase diagram with many features in common with those of other models, except that  incorporation of the last subunit was hindered by the internal degrees of freedom of the nearly complete partial capsids (see below).

Doye, Louis, and coworkers \cite{Wilber2009a,Johnston2010,Wilber2009,Wilber2007} used Monte Carlo simulations to study the dynamics and thermodynamics of a variety of patchy-sphere models, with ground state geometries that include the set of regular polyhedra, $T$=1 shells, and $T$=3 shells. For interaction potentials which did not incorporate dihedral angles (motivated by patchy colloids), they found that assembly could proceed through a disordered liquid state intermediate for some  parameter ranges \cite{Wilber2007} and competition with disordered states led to a dodecahedra being kinetically inaccessible \cite{Wilber2009}. The liquid state disappeared when the dihedral potentials consistent with protein-protein interactions were included.

 Hicks and Henley \cite{Hicks2006} proposed a model for assembly of HIV capsids in which triangular subunits were irreversibly attached to sites on a growing cluster according to the local geometry to form hexagonal or pentagonal substructures, with a subunit-subunit interaction geometry that defined a preferred spontaneous curvature. They found that under irreversible attachment the model produced an ensemble of irregular structures, but not the conical shells observed in EM images of mature HIV capsids \cite{Ganser-Pornillos2004,Benjamin2005}. Levandovsky and Zandi \cite{Levandovsky2009} extended and modified the model to allow merging and for the structure to minimize its elastic energy at each  step. The model predicted an ensemble of structures which closely resemble those seen in retrovirus capsids, and suggested that the protein spontaneous curvature plays a key role in determining the capsid shape (i.e. spherical, conical, etc.).

Higher resolution representations of capsid proteins have been used to simulate parts of assembly pathways. For example Chen and Tyco \cite{Chen2011} developed a model for the HIV capsid protein which includes information about subunit-subunit contacts derived from NMR studies, and simulated the  early stages of assembly in two dimensions. Tunbridge et al. \cite{Tunbridge2010} modeled the assembly of small oligomers of HBV proteins with the proteins modeled as rigid bodies using a transferable one-bead-per-amino acid model \cite{Kim2008}.  Futhermore systematic coarse graining from atomistic simulations was used to estimate subunit positions and orientations from cryo electron microscopy images of the immature HIV capsid \cite{Ayton2010}. A study of dimerization of the C-terminal domain of the HIV capsid protein  at atomic resolution with explicit water showed that water in the vicinity of the protein-protein interface sits at the edge of a drying transition \cite{Yu2012}.

\subsection{Conclusions from assembly dynamics models}
\label{sec:assemblyDynamicsConclusions}

Having presented an overview of some of the theoretical and computational models of capsid assembly dynamics, we highlight a few of the more important conclusions that have emerged these studies. We will see that many of the predictions are  consistent across both rate equation and particle-based models and match experimental observations.

\begin{figure} [bt]
\begin{center}
\epsfig{file=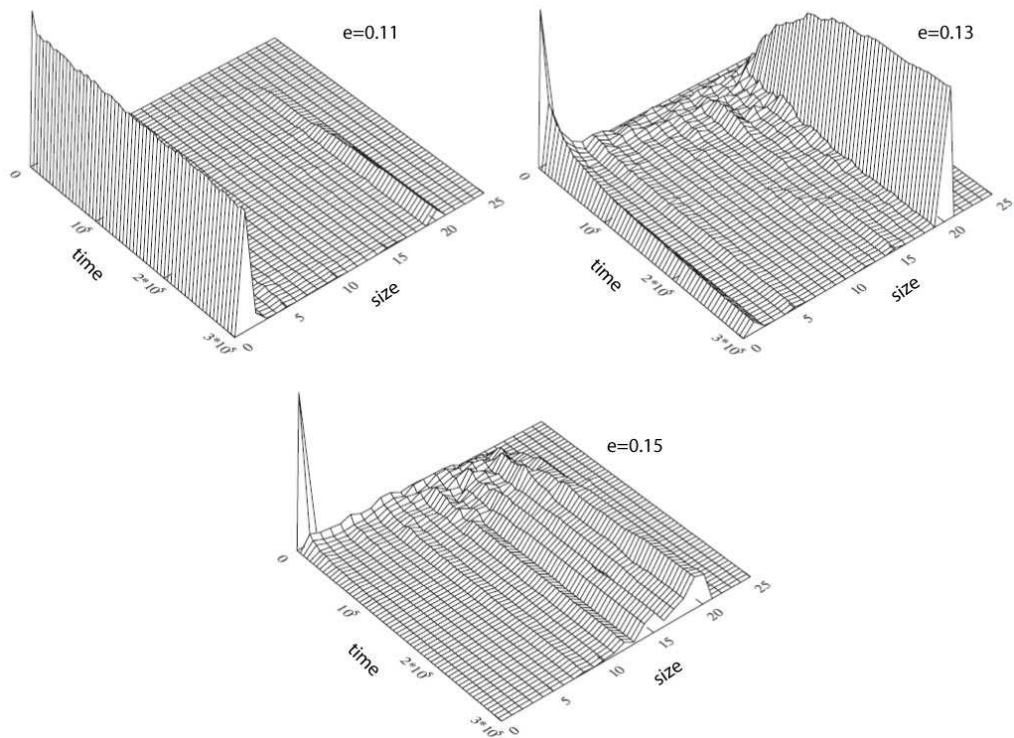,width=.75\textwidth}
\caption{\label{fig:RapaportFigure}
The time evolution of cluster size distributions are shown for three interaction strengths, parameterized by $e$, for molecular dynamics simulations of the triangular subunit model shown in Fig. \ref{fig:computationalModels}B. The model capsid comprises 20 subunits; the system has entered a kinetic trap at the highest interaction strength. Figure reprinted with permission  from Phys. Biol., {\bf 7}, 045001 (2010), Rapaport, \emph{Modeling capsid self-assembly: design and analysis}, Copyright (2010) IOP Publishing.
}
\end{center}
\end{figure}

{\bf Capsid assembly kinetics are sigmoidal.}
Consistent with the experimental measurements, the theoretical and computational models predict sigmoidal assembly kinetics.  Example predictions are shown for rate equation models in Figs. \ref{fig:lightScatter}B and \ref{fig:KumarRateEquationsFitToData} for Brownian dynamics simulations in  Fig.  \ref{fig:lagTimes}A and for molecular dynamics simulations in Fig. \ref{fig:RapaportFigure}. In all cases,  there is an initial lag phase during which capsid intermediates form, followed rapid capsid production and then an asymptotic approach to equilibrium.

{\bf Intermediates do not build up.} Fig. \ref{fig:RapaportFigure} shows the fraction of subunits in intermediates of all sizes as a function of time for a model icosahedron \cite{Rapaport2010} for three values of the binding energy. For the two smaller values which lead to productive assembly there is never a significant fraction of subunits found in intermediates. A similar result is found for both rate equation models and other computational models, consistent with experiments where intermediates are generally not detectable for productive parameters.

 {\bf The duration of the lag phase is set by the mean capsid assembly time.} This observation was described in section \ref{sec:lagTimes} (Fig. \ref{fig:lagTimes}B).

\begin{figure} [bt]
\begin{center}
\epsfig{file=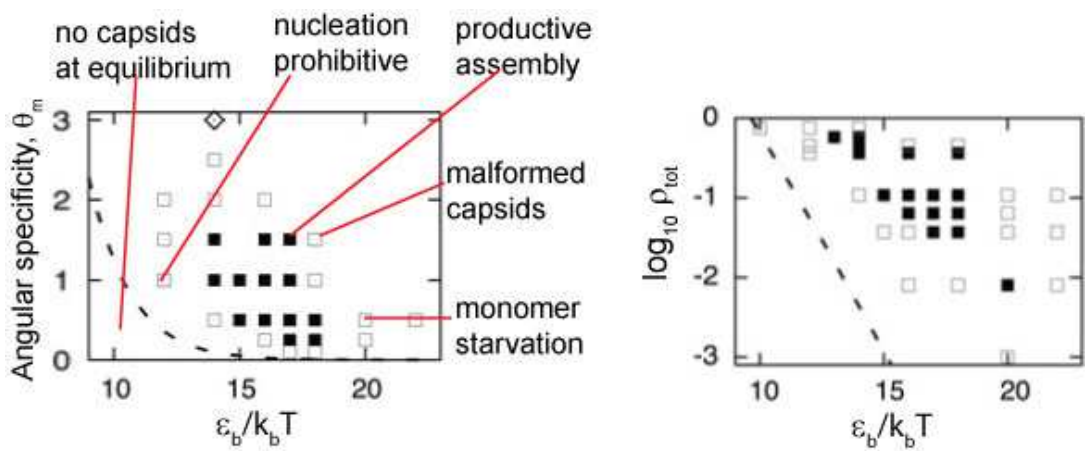,width=.8\textwidth}
\caption{ Assembly products at long times for 60-subunit patchy-sphere model $T$=1 shells as functions of binding energy $\eb$, angular specificity, $\theta_\text{m}$, and total particle concentration $\rhoTot$. Solid squares indicate parameter sets for which there were significant yields of well-formed capsids $\fc\ge0.3$, while open squares indicate poor assembly, $\fc<0.3$. The dashed line indicates the parameter values above which significant capsid assembly should occur at equilibrium. The location of the five regimes discussed in the text are shown on the phase diagram on the left. Figure based on Ref. \cite{Hagan2006}. }
\label{fig:simulationPhaseDiagramsA}
\end{center}
\end{figure}

\begin{figure} [bt]
\begin{center}
\epsfig{file=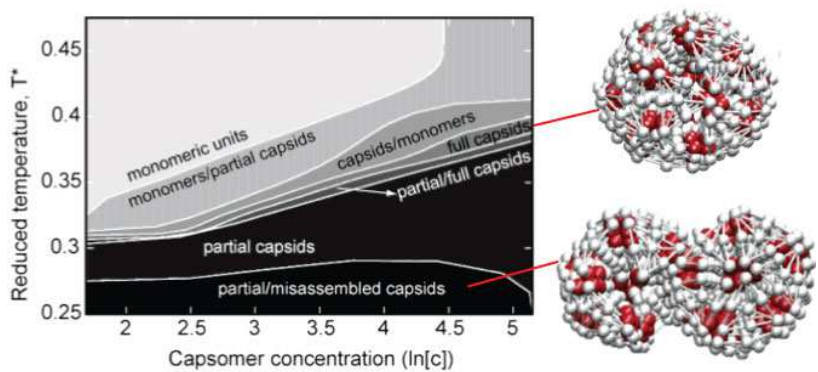,width=.6\textwidth}
 \caption{ Assembly products at long times for a 20-subunit extended subunit $T$=1 shell as a function of temperature (i.e. inverse of interaction strength) and particle concentration. Representative structures are shown for the well-formed and mis-assembled regions.  Figure adapted with permission from  Nano Lett., {\bf 7}, 338-344 (2007), \emph{Deciphering the kinetic mechanism of spontaneous self-assembly of icosahedral capsids}, Nguyen, Reddy, and Brooks, Copyright (2007) American Chemical Society. }
  \label{fig:simulationPhaseDiagramsB}
\end{center}
\end{figure}

\begin{figure} [bt]
\begin{center}
\epsfig{file=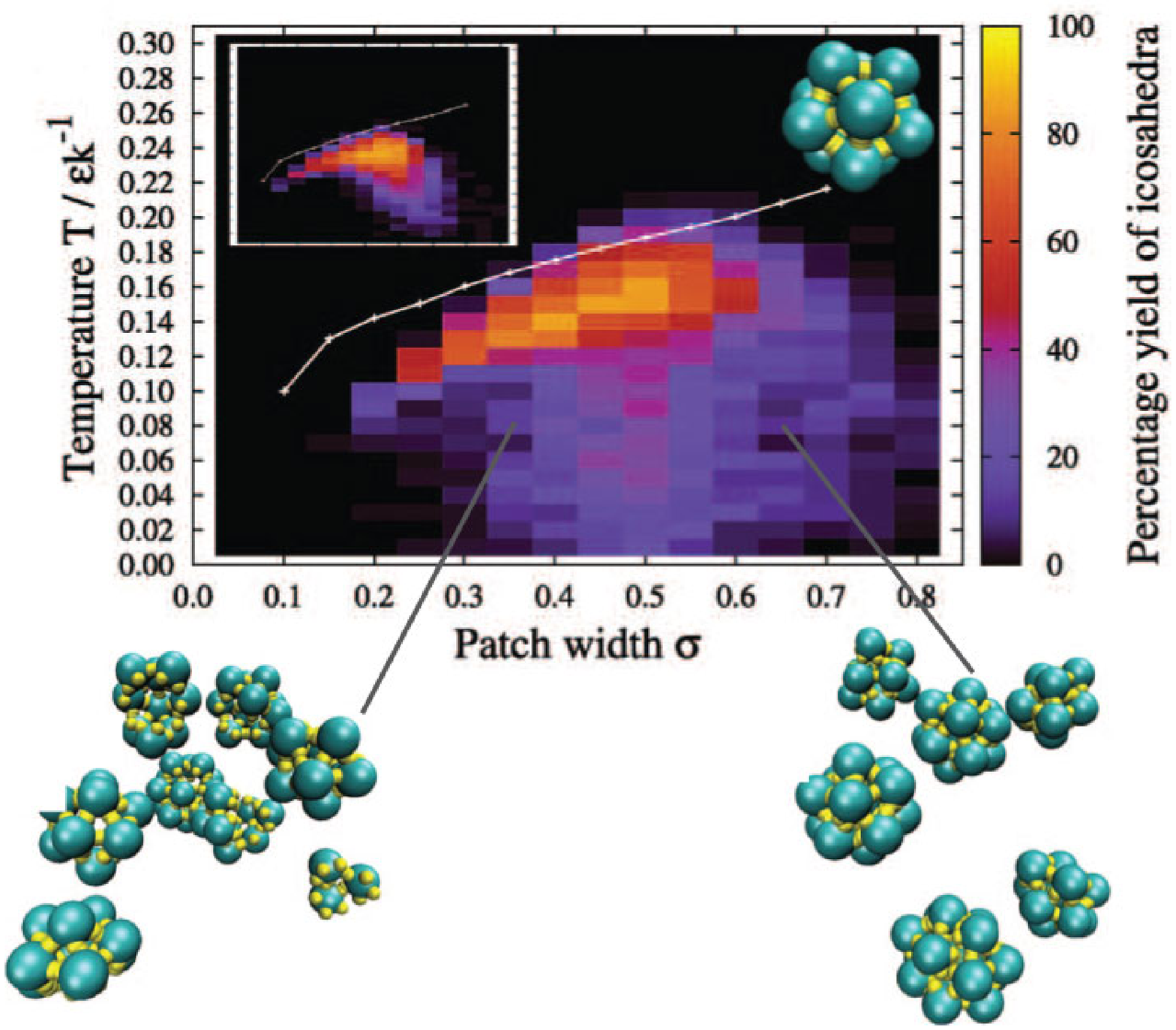,width=.43\textwidth}
 \caption{ Assembly products at long times for a 12-subunit patchy-sphere model icosahedron. The fraction of subunits in  target clusters is shown as a function of the patch width $\sigma$ (measured in radians) and
reduced temperature. The inset shows the equivalent plot for a system with the same parameters except without dihedral terms in the interaction potential. The image at the top right shows the target structure, while the lower images show regions of the system for simulation at the indicated parameter values.
The white lines show the temperature for the equilibrium
transition from assembled clusters to a gas of monomers  calculated from umbrella sampling. Figure and images reprinted with permission from J. Chem. Phys., {\bf 131}, 175102 (2009), \emph{Monodisperse self-assembly in a model with protein-like interactions}, Wilber, Doye, Louis, and Lewis, Copyright(2009) American Institute of Physics.}
\label{fig:simulationPhaseDiagramsC}
\end{center}
\end{figure}

 {\bf Optimal assembly occurs when subunit-subunit association is reversible.}
Dynamical simulations predict that capsid yields at finite observation times are nonmonotonic with respect to control parameter values (e.g. subunit concentration, binding energy, or specificity), which is consistent with experimental observations \cite{Zlotnick1999,Zlotnick2000,Ceres2002}. The plots showing yield as a function of time in Figs. \ref{fig:lightScatter}B, \ref{fig:lagTimes}A, and \ref{fig:RapaportFigure} demonstrate suppressed capsid production at the highest subunit concentration or binding energy. Similarly, examples for which dynamical simulation results at long times have been presented in cross-sections of phase diagrams are shown in Figs. \ref{fig:simulationPhaseDiagramsA}, \ref{fig:simulationPhaseDiagramsB} and \ref{fig:simulationPhaseDiagramsC}. \footnote{In these `kinetic phase diagrams', a timescale is chosen for which the results at the interesting regions of parameter space have stopped changing on timescales which are long compared to those accessible to simulations. As noted in section \ref{sec:approachEquil}, assembly reactions approach equilibrium asymptotically and thus it is not possible to make a true equilibrium observation from a dynamical simulation started from unassembled subunits. However, for much of parameter space the system reaches an outcome which is nearly stationary with respect to experimentally relevant timescales due to the fact that bond-breaking is an activated process.}.

It is worth noting that the subunit-subunit binding free energy ($\gb$ in \eqref{eq:Gi}) cannot be directly determined from force field parameters in the simulations referenced in Figs.~\ref{fig:simulationPhaseDiagramsA}, \ref{fig:simulationPhaseDiagramsB} and \ref{fig:simulationPhaseDiagramsC}, as the binding entropy penalty depends on the length scale and directional specificity of the interaction. Binding free energies estimated from umbrella sampling (e.g. \cite{Hagan2006,Wilber2009,Hagan2011}) for optimal parameter values were of order $5-10 \kt$ depending on particle concentrations, consistent with experimental observations (see Fig. \ref{fig:fExpVsPhi} \cite{Ceres2002}).

\begin{figure}
\begin{center}
\includegraphics[width=0.5\textwidth]{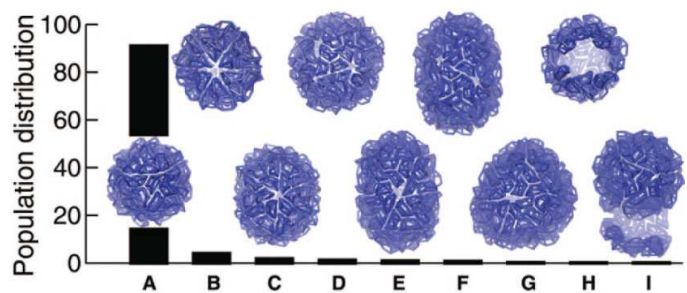}
\caption{Population distribution of structures obtained at long times for near optimal parameters using discontinuous molecular dynamics for a $T$=1 model by Nguyen et al. \cite{Nguyen2009}. The structures were defined by Nguyen et al. \cite{Nguyen2009} as (A) complete icosahedral capsids, (B) oblate capsules, (C) angular
capsules, (D) twisted capsules, (E) tubular capsules, (F) prolate capsules, (G)
conical capsules, (H) partial capsids, and (I) open mis-aggregates. Figure reprinted with permission from J. Am. Chem. Soc. {\bf 131}, 2606-14 (2009), \emph{Invariant polymorphism in virus capsid assembly}, Nguyen, Reddy, and Brooks, Copyright(2009) by the American Chemical Society.}
\label{fig:nguyenMalformed}
\end{center}
\end{figure}

The phase behavior can be separated into five regimes, whose locations are indicated on the phase diagram shown in Fig.~\ref{fig:simulationPhaseDiagramsA}:

\begin{enumerate}
\item \emph{No assembly at equilibrium.} In this regime the interactions driving assembly are too weak to overcome the rotational and mixing entropy of free subunits, and virtually all subunits are present as free dimer in equilibrium.  As discussed in section \ref{sec:thermo}, this regime corresponds to  subunit concentrations below a critical value $\rho < \rhoStar$, whose value depends on the subunit-subunit binding free energy; the values of $\rhoStar$ are shown as dashed and white lines in Figs.~\ref{fig:simulationPhaseDiagramsA} and  \ref{fig:simulationPhaseDiagramsC} respectively.

\item \emph{No assembly on relevant time scales due to a nucleation barrier.} For concentrations sufficiently close to the critical value, $\rho\gtrsim\rhoStar$, nucleation barriers are prohibitive (see section \ref{sec:approachEquil}and Fig. \ref{fig:nucleationBarriers}). Thus assembly is not seen on timescales that are accessible to simulations (or experiments) at these concentrations. This is the first kinetic effect that can prevent assembly at long but finite times. This regime is seen in the phase diagrams shown in Figs.~\ref{fig:simulationPhaseDiagramsA} and \ref{fig:simulationPhaseDiagramsC}, where there is a region between $\rhoStar$ and parameter values at which assembly is observed.

\item \emph{Productive assembly.} For moderate parameter values initial nucleation barriers are small enough such that finite-time assembly yields can be quite large, with $\fc \gtrsim 90\%$.

\item \emph{Free subunit starvation kinetic trap.}
The first form of kinetic trap described in section \ref{sec:assemblyTimeScales} arises due to the constraint of mass conservation.  When nucleation is fast compared to elongation ($\ckt$ in Eq.~\ref{eq:ckt}), too many capsids nucleate at early times, and the pool of free subunits or small intermediates becomes depleted before a significant number of capsids are completed.
 This phenomenon can be seen readily in the time series shown in the right panel of Fig. \ref{fig:RapaportFigure}.
 Except to the extent that the remaining partial capsids have geometries which allow for direct binding, further assembly requires the analog of Ostwald ripening, in which subunits unbind from smaller partial capsids and are scavenged by larger intermediates.
 This is an activated process since it requires bond-breaking; hence it is generally slow in comparison to time scales for assembly at parameter sets that do not lead to trapping. This form of kinetic trap was first predicted with rate equations models by Zlotnick and coworkers \cite{Zlotnick1999,Endres2002}, and was shown to be consistent with experiments (e.g. the largest salt concentration in Fig. \ref{fig:lightScatter}A).

\item \emph{Malformed capsids.} The second form of kinetic trap  arises when subunits forming strained bonds that deviate from the ground state of the interaction potential are trapped into growing clusters by subsequent subunit additions. For example, in $T$=1 capsids it is common to observe hexameric defects at the  five-fold vertices. In some cases defects lead to closed shells that lack icosahedral symmetry, whereas in other cases the curvature is disrupted significantly enough that spiral structures form. Nguyen and coworkers \cite{Nguyen2009} catalogued sets of hexameric defects that lead to closed or open structures  (see Fig. \ref{fig:nguyenMalformed}).
\end{enumerate}

 The predictions and observations that capsid assembly yields are nonmonotonic with respect to driving forces have contributed to a wider understanding that many examples of self-assembly are most efficient when structures are stabilized by numerous relatively weak interactions  \cite{Whitesides2002,Ceres2002,Hagan2006,Jack2007,Zlotnick2007,Rapaport2008,Elrad2008,Whitelam2009,Nguyen2007,Wilber2007,Wilber2009,Klotsa2011,Grant2011,Grant2012}.  While strong interparticle bonds stabilize the ordered equilibrium state, they also promote and stabilize the two kinetic traps described above that frustrate assembly. Thus, effective self-assembly proceeds by relatively transient bond formation, with  bond-breaking events that are nearly as frequent as bond formations. This idea was first suggested in the context of virus assembly by Zlotnick and coworkers \cite{Zlotnick1999,Ceres2002} ( Fig.~\ref{fig:fExpVsPhi}) and by Schwartz and coworkers \cite{Schwartz1998} based on preliminary particle-based simulations. The extent to which capsid assembly reactions approach reversibility has been monitored by a variety of metrics (e.g. \cite{Jack2007,Rapaport2008,Rapaport2010,Hagan2011}) which include measuring relative frequencies of bond formation and bond-breaking \cite{Jack2007,Rapaport2008,Rapaport2010}, fluctuation dissipation ratios \cite{Jack2007}, and the extent to which clusters of similar size  are in relative Boltzmann equilibrium \cite{Hagan2011}. It has been  shown that the extent to which reversibility is violated can be correlated to yields of well formed capsids. Similar approaches have been applied to models for crystallization ( e.g. \cite{Jack2007,Hagan2011,Grant2011,Grant2012,Klotsa2011}). Despite the fact that viral capsids are monodisperse closed shells whereas crystals are extended structures, the correlations between reversibility and assembly yields in models of crystallization are strikingly similar to those observed for models of capsid assembly.

Given the significant amount of attention which has been accorded to reversibility, it is important to note that the strong interactions (large degree of supersaturation) which lead to violations of reversibility contribute to assembly failure through both of the kinetic traps described above. In the case of the free subunit starvation trap, the Boltzmann factor in Eq. \ref{eq:ckt} identifies that nucleation rates increase much more quickly than elongation rates and there is a threshold binding energy or subunit supersaturation above which the condition will be violated. Furthermore, once free subunit starvation sets in, the Ostwald ripening which eventually leads to equilibration requires bond-breaking and hence is characterized by a timescale which increases exponentially with binding energy.

The malformed capsid trap arises when subunit addition to a growing partial capsid occurs more quickly than already associated subunits can anneal defective or strained interactions. The rate of subunit addition is proportional to free subunit concentration. Annealing in general requires some bond-breaking and hence is characterized by a timescale which increases exponentially with binding energy.

While both the free subunit starvation trap and malformed capsids arise for strong interactions or high subunit concentrations, which effect dominates at a given parameter set is a strong function of the directional specificity. Highly specific binding interactions imply that imperfect subunit interactions are unstable and only very strong binding energies lead to malformed capsids. Therefore, for interactions with sufficiently high directional specificity, the threshold binding energy for malformed capsids is well above the threshold binding energy for  the free subunit starvation trap (Eq. \ref{eq:ckt}).
From an evolutionary standpoint, or from the perspective of designing synthetic or biomimetic assembly systems,  it might appear that maximizing directional specificity would be optimal for productive assembly. However, increasing directional specificity reduces the subunit kinetic cross-section and thus leads to lower subunit-subunit binding rates  \cite{Hagan2006,Whitelam2009}. There is a trade-off between selectivity and kinetic accessibility  which limits the optimal degree of directional specificity for finite-time assembly reactions. In the case of capsid proteins the degree of directional specificity which is physically realizable is limited by the nanometer length scale of the underlying hydrophobic and electrostatic interactions. The degree of specificity used within coarse-grained models can be roughly estimated based on the physical interaction length scales (e.g. \cite{Nguyen2007}) or it can be calculated systematically via coarse graining \cite{Ayton2010,Hicks2010}.

{\bf Experimental observations of partial or malformed capsids.}
While light scattering experiments on HBV \cite{Zlotnick1999} and CCMV \cite{Zlotnick2000} clearly indicate that stronger than optimal interactions lead to reduced assembly yields, these results were interpreted in the context of the monomer starvation trap. In general products of reactions performed at stronger than optimal parameters have not been well characterized by imaging, in part because variable sizes and defective structures  limit the use of multi-particle and/or icosahedral averaging, but also because effort has been focused on parameter sets which produce high yields of capsids. However in a notable early \emph{in vitro} study Sorger et al. \cite{Sorger1986} identified malformed turnip crinkle mosaic virus capsids assembling around the  genomic RNA. Furthermore, Teschke and coworkers \cite{Parent2007a} have catalogued a series of malformed structures which result due to mutations in the P22 capsid protein, and Stray et al.  \cite{Stray2005} showed that HBV assembly is accelerated and defective in the presence of an antiviral drug.

%from JCP2011:
%We note that there are two mechanisms by which cluster equilibration can be violated. In the first, subunits form strong interactions with a sub-optimal number of partners. In other words, each subunit-subunit interaction approximately corresponds to a minimum in the interaction potential, but subunits do not add on to a growing cluster in locations that offer the most interaction partners.
%In the second mode of violation, subunits form strained bonds which deviate from the ground state of the interaction potential. For example, assembling capsids frequently form hexameric defects, as illustrated in   Fig.~\ref{fig:caps_mallet}. The first form of cluster equilibration violation can be incorporated into the rate equation approach, at a cost of significantly increased computational complexity, if the space of all possible cluster configurations can be predefined, and then relevant cluster configurations can be enumerated ahead of time \cite{Moisant2010} or sampled stochastically \cite{Sweeney2008}. However, these approaches have not been used to address the possibility of defective bonds, for which it is not possible to predefine the space of possible cluster configurations.

%\subsubsection{The role of solvent}
%\label{sec:roleOfSolvent}
{\bf The role of solvent.}
The solvent plays several important roles in capsid assembly. As described in section \ref{sec:drivingForces}, subunit-subunit association is typically driven by hydrophobic interactions with moderate ionic strengths required to screen electrostatic interactions.
   These solvent-mediated effects have been incorporated implicitly in particle-based capsid assembly simulations via the subunit-subunit interaction potentials. Second, the solvent absorbs kinetic energy released by the formation of low-energy subunit-subunit interactions. Third, random buffeting by the solvent helps annealing. The latter two effects have been incorporated through a hydrodynamic drag and stochastic buffeting force in Brownian dynamics \cite{Schwartz1998,Hagan2006,Elrad2008,Elrad2010,Hagan2011} or through explicit inclusion of inert solvent particles by Rapaport \cite{Rapaport2008,Rapaport2010}. Analysis of simulated trajectories from both approaches indicated that the random buffeting plays an important role in annealing by inducing dissociation of improperly bound subunits. In contrast, simulations which used a thermostat but no implicit solvent \cite{Rapaport2004} demonstrated much less efficient annealing during assembly because improperly bound subunits dissociated only upon collision with another subunit.

A fourth effect of solvent is to introduce hydrodynamic interactions between subunits. This effect is typically not accounted for within a Brownian dynamics simulation because of the computational expense \cite{Schlick2002}. Hydrodynamic coupling has been effectively included in assembly simulations using a cluster move Monte Carlo \cite{Whitelam2011} and is intrinsically included in the explicit solvent simulations \cite{Rapaport2008,Rapaport2010}, albeit  at a significant computational expense.  The primary effect of neglecting hydrodynamic interactions is that bonded clusters are `freely draining', and thus experience a hydrodynamic drag that is proportional to the number of subunits rather than  the cluster hydrodynamic radius.  Hydrodynamic interactions also influence subunit collision rates, but a quantitative estimate of collision rates is likely to require atomic-resolution models in any case. To this point, no significant differences in assembly behavior between explicit solvent and Brownian dynamics simulations have been  reported.

\subsection{Differences among models}
\label{sec:differences}
The models described in this section are highly simplified and thus designed to uncover generic assembly mechanisms for icosahedral shells. In that regard it is encouraging that results described above are consistent across most or all of the models. We now discuss some of the key differences  between the models.

We already noted in the previous section that state-based models have not been able to describe the malformed capsid trap. Among particle-based models,  the structural details of ensembles of malformed structures predicted by dynamical simulations are somewhat dependent on details of the model interaction geometries. E.g., trapezoidal subunits can form defects not available to triangular subunits and thus do not give rise to the same ensemble monster structures \cite{Nguyen2007,Nguyen2009}.  Similarly, we have found that patchy sphere models can form closed shells which are smaller than the ground state geometry under strong interactions, which does not seem to occur for triangular or trapezoidal subunits. Also, Ref. \cite{Hagan2006} found that different subunit valencies can lead to different  sets of assembly pathways even if the overall assembly efficiency and kinetics are similar.

{\bf Insertion of the last subunit.} In all particle-based models insertion rates decrease as the capsid nears completion due to steric hindrances. This effect appears to be amplified in the model of Nguyen et al. \cite{Nguyen2007,Nguyen2009}, where insertion of the final subunit suppresses internal vibration modes of the partial capsid. Insertion of the final subunit was found to be rate limiting and even free energetically unfavorable for some parameter ranges. Experimental evidence supporting this observation has been lacking; for example mass spectrometry studies of HBV capsids indicated an overwhelming preponderance of complete capsids \cite{Uetrecht2008,Uetrecht2010a}. It is worth noting though that most imaging studies of capsid structures use icosahedral and/or multiparticle averaging and thus would not detect missing subunits. Although experimetally challenging, a systematic search for defects, including missing subunits, under a variety of assembly conditions would be of great interest.

\subsection{Higher $T$ numbers}
\label{sec:higherTNumber}
As described in section \ref{sec:anatomies}, icosahedral capsids with more than 60 proteins ($T>1$) require precise arrangements of pentameric and hexameric capsomers, often with different protein conformations in the distinct but `quasi-equivalent' sites. We first present studies that investigated the relative thermodynamic stabilities of possible capsid structures, followed by studies that attempted to discover how  the correct arrangement of conformations (or of pentamers and hexamers) is achieved during the dynamics of shell construction.

\subsubsection{Structural stability of different capsid geometries}
\label{sec:structuralStability}
The relative thermodynamic stabilities of shell geometries have been investigated by Monte Carlo simulations.  Zandi et al. \cite{Bruinsma2003,Zandi2004} studied shells comprised of two species of discs,  representing pentamers and hexamers confined to a spherical surface, and found that the Caspar and Klug geometries correspond to minimum free energy configurations for appropriate size-ratios of the discs. Chen, Zhang, and Glotzer \cite{Chen2007} studied the assembly thermodynamics of cone-shaped particles. For decreasing cone angles the particles assembled into convex shells of increasing sizes corresponding to `magic numbers'. Certain magic numbers corresponded to icosahedral shells, and the assembled structures were found to correspond to  equilibrium structures of colloids confined to spherical surfaces. Similar approaches have been used to investigate the relative stabilities of potential prolate structures. Chen et al. \cite{Chen2007a} performed Monte Carlo simulations of spheres packing on prolate surfaces and found structures consistent with some prolate virus capsids, while Luque, Zandi and Reguera \cite{Luque2010} found that free-energy minimization of disks representing pentamers and hexamers confined to a prolate surface led to structures consistent with a number of prolate or bacilliform capsid structures and identified selection rules for the length, structure, and number of capsomers for prolate capsids.

Fejer, Chakrabarti, and Wales \cite{Fejer2010} studied the formation of shells by disk-shaped or ellipsoidal subunits with anisotropic interactions that dictated a preferred curvature, via searching for minima in the configurational energy landscape. For disk-shaped subunits the results resembled those of the  assembling cones \cite{Chen2007} with global minima corresponding to icosahedral structures for particular preferred curvatures. For appropriate arrangements of anisotropic interactions, they were also able to describe tubular, helical and head-tail morphologies (i.e. resembling tailed bacteriophages), as well as multi-shelled structures.

Continuum elasticity theory has also shed light on capsid shapes. The fact that small icosahedral capsids tend to be spherical while larger capsids tend to look more faceted (i.e. icosahedral) \cite{Baker1999} was reproduced by continuum elastic models in which the faceting corresponds to a buckling transition \cite{Lidmar2003,Siber2006}. A similar approach was applied to spherocylindrical and conical shells \cite{Nguyen2005} and reproduced features of retrovirus capsid shapes \cite{Nguyen2006}.

\subsubsection{Dynamics of forming icosahedral geometries}
\label{sec:largeTDynamics}
Berger et al. proposed a system of `local rules' \cite{Berger1994} in which subunit-subunit binding interactions are highly conformation specific. I.e., subunits with a particular conformation, A, bind strongly to A-sites on a growing capsid (binding sites for which the conformations of neighboring subunits favor the A conformation), but bind weakly or not at all to other sites. Simulations were performed using trivalent subunits, with different subunit conformations represented by differences in interaction geometries. Assembly dynamics was modeled by placing subunits one-at-a-time at binding sites on a growing capsid according to the local rules. It was found that icosahedral capsids could assemble with high fidelity, even under a certain degree of flexibility in subunit-subunit interaction angles. Relaxation of local rules led to the formation of malformed structures, such as spiraling shells \cite{Schwartz2000} and one of the first suggestions that it might be possible to develop antiviral agents that act by misdirecting capsid assembly \cite{Prevelige1998}. The local rules are generalizable and were used to describe papovirus (\emph{polyomaviridae}) assembly \cite{Schwartz2000}.

Several of the molecular dynamics studies of capsid assembly described above modeled assembly into $T$=3 or $T$=4 capsids \cite{Rapaport2004,Nguyen2009,Elrad2008,Johnston2010,Rapaport2010} and Nguyen and coworkers \cite{Nguyen2008} modeled the assembly of preconstructed pentamers and hexamers into capsids as large as $T$=19. In most cases these studies used conformation-dependent subunit-subunit interaction energies corresponding to strict local rules, with each subunit locked into a particular conformation. They found that larger capsids can assemble under conditions similar to those which lead to well-formed $T$=1 capsids and are constrained by similar kinetic traps. However, in general parameters need to be more tightly tuned the larger the capsid is, as there are more opportunities for misdirection of the assembly pathway and elongation times are longer (hence requiring longer nucleation times according to Eq. \ref{eq:ckt}). As noted above, the ensemble of malformed structures that results for non-optimal parameters is somewhat dependent on subunit interaction geometries, and thus depends on the preferred $T$-number.  Ref. \cite{Nguyen2009} also found that details of assembly mechanisms  could differ between $T$=1 and $T$=3 capsids.

{\bf Breaking the local rules.} There are two experimental observations which seem inconsistent with the assumption of strict local rules, at least for small icosahedral viruses. First, for many viruses the structures of binding interfaces are quite similar for subunits in different conformations (see e.g. Ref. \cite{Tang2006}). Second, capsid proteins which assemble into a particular icosahedral structure with high fidelity can adapt to form different icosahedral morphologies under different conditions \cite{Bancroft1970,Adolph1974,Lavelle2009}. Furthermore, as described further in section \ref{sec:cargo}, capsids exhibit polymorphism when assembling around cargoes with incommensurate sizes.  For example, Dragnea and coworkers \cite{Sun2007, Dixit2006, Chen2005, Dragnea2003} demonstrated that brome mosaic virus (BMV) proteins assemble into  $T$=1, pseudo-T2, and $T$=3  capsids around charge-functionalized nanoparticles with different diameters. These observations of polymorphism in empty and cargo-containing capsids raise the questions of how strongly subunit-subunit interactions across a given interface can depend on conformation (i.e., how strict are the local rules), and how strong of a conformation-dependence is required to assemble into icosahedral structures.

Elrad and Hagan simulated assembly of $T$=3 empty capsids with a model in which the conformation-dependence of the subunit-subunit interaction energies was systematically varied. They found that $T$=3 capsids could form with high fidelity provided that interactions which violate the conformation-dependence of the $T$=3 structure were 20\% weaker than those consistent with the target geometry. If the conformation dependence of the interactions was less specific, asymmetric closed shells or open spiraling structures were the dominant assembly morphologies. Interestingly, since the binding free energy at which capsids assembled successfully is only on the order of $5-10 \kt$, a 20 \% reduction in free energy for subunits with the wrong set of conformations corresponds to a free energy difference only on the order of $\kt$, which could easily arise from minor variations in binding interfaces. It was found that this level of conformation-dependence did allow for adaptable assembly into alternative icosahedral geometries around nanoparticles with different sizes \cite{Elrad2008}.

The difficulty of dynamically constructing icosahedral shells in the absence of conformation-specific interactions is elucidated by the recent study of Luque et al. \cite{Luque2012}, which examined the dynamics of monolayers of spheres constrained to growing on a spherical manifold. Recall that a series of `magic number' equilibrium structures were identified for this  system in Ref.  \cite{Chen2007}. Luque et al. \cite{Luque2012} found that line tension of the growing shell drives premature closure of the shell thus hindering formation of defect-free structures. Other mechanisms promoting disorder became important for structures comprised of more than 50 subunits. Interestingly, small defect-free shells could be assembled by adjusting the radius of the manifold. The important roles that  template can play in directing assembly will be discussed in section \ref{sec:cargo}.

Recent studies indicate that RNA-protein interactions play a key role in determining assembly pathways for the ssRNA bacteriophage MS2 \cite{Stockley2007,Rolfsson2008,Basnak2010,Morton2010,Dykeman2011,ElSawy2010,Dykeman2010,Dykeman2010a,Toropova2008}. Stockley and coworkers used mass spectrometry to show that conformational switching of the MS2 capsid protein can be regulated by binding of short RNA step-loops, with sequence-dependent activity. Dykeman and Twarock performed all-atom normal modes calculations on the capsid protein in the presence and absence of RNA which identified a potential allosteric connection between the RNA binding site and the flexible FG loop which undergoes the bulk of the conformational change \cite{Dykeman2010a}. Coarse-grained computational modeling indicated that RNA binding influences subunit-subunit association rates and a conformation-dependent manner \cite{ElSawy2010}. Furthermore, mass spectrometry identified two intermediates in the assembly pathway \cite{Stockley2007} as well as their concentrations during assembly. This information was used to build and fit parameters for a Master equation model of assembly \cite{Morton2010}. The results indicated that there are two dominant assembly pathways, whose prevalence is determined by the stoichiometric ratios of RNA and capsid protein.

\section{Cargo-containing capsids}
\label{sec:cargo}
\subsection{Structures}
\label{sec:cargoExperiments}
In this section we consider capsid assembly around RNA or other types of cargo. The  structures of numerous virus capsids assembled around single-stranded nucleic acids have been revealed to atomic resolution by x-ray crystallography and/or cryo-electron microscopy (cryo-EM) images  (e.g.\cite{Fox1994,Valegard1997,Johnson2004,Tihova2004,Krol1999,Stockley2007,Toropova2008,Lucas2002,Valegard1994,Worm1998,Grahn2001,Valegard1997,Helgstrand2002,Schneemann2006}).
The packaged nucleic acids are less ordered than their protein containers and hence have been more difficult to characterize. However cryo-EM experiments have identified that the nucleotide densities are nonuniform, with a peak near the inner capsid surface and relatively low densities in the interior\cite{Tihova2004,Zlotnick1997,Conway1999}. For some viruses striking image reconstructions show that the packaged RNA adopts the symmetry of its protein capsid (e.g. \cite{Tihova2004,Toropova2008,Schneemann2006}). While atomistic detail has not been possible in these experiments, all-atom models have been derived from equilibrium simulations \cite{Freddolino2006,Devkota2009}.

\subsection{The thermodynamics of core-controlled assembly}
\label{sec:cargoThermo}
In this section we extend the thermodynamic analysis of section \ref{sec:thermo} to include interactions with an attractive core, following the calculations of Zandi and van der Schoot \cite{Zandi2009} and Hagan \cite{Hagan2009}.
We consider a dilute solution of capsid protein subunits with total density $\rhoTot$, and cores (i.e. polymers or nanoparticles) with density $\coreTot$.  We define a stoichiometric ratio as the ratio of available cores to the maximum number of capsids which can be assembled, $r=N \coreTot/\rhoTot$, with $N$ the number of subunits in a complete capsid. Subunits can associate to form capsid intermediates in bulk solution or on core surfaces.

Following section \ref{sec:thermo} we minimize the total free energy under the constraints that the total subunit and core concentrations are fixed:
\begin{align}
 \rhoTot & = \sum_{n=1}^N n(\rho_n + x_n)\nonumber \\
 \coreTot&= \sum_{n=0}^N x_n
\label{eq:CC}
\end{align}
with $\rho_n$ and $x_n$ respectively the concentrations of empty capsid intermediates with $n$ subunits or cores complexed with $n$ subunits (for simplicity we assume here that cores cannot form complexes with more than $N$ subunits and that $x_N$ gives the concentration of well-formed capsids assembled around cores).  We then arrive at two laws of mass action (compare to Eq. \ref{eq:rhon})
\begin{align}
\rho_n v_0&=\left(\rho_1 v_0\right)^n\exp[-\beta \Gcap_n] \nonumber \\
x_n v_0 &=v_0 x_{0} \left(\rho_1 v_0\right)^{n}\exp[-\beta \Gcore_n]
\label{eq:Zcore}
\end{align}
with $x_0$ the concentration of empty cores. Here $\Gcap$ is the empty capsid free energy (Eq. \ref{eq:Fmt}) discussed in section \ref{sec:thermo}.  The quantity  $\Gcore_n$ is the free energy for a core with $n$ subunits which includes the core-subunit  interactions. These are the crucial interactions that drive capsid proteins to assemble around the core, and models for this quantity are discussed next.

 As shown in section \ref{sec:thermo} there is a threshold concentration $\rhoStar=\exp[\beta \Gcap_N/(N-1)]$ below which essentially no empty capsid assembly occurs (Eq.~\ref{eq:fc}). However, the core-subunit interactions can further stabilize a complete capsid so that $\Gcore_N<\Gcap_N$ and core-assisted assembly can occur at lower concentrations. This capability is exploited by  many ssRNA viruses, whose capsids assemble only in the presence of RNA or other polyanions at physiological conditions.

For core-controlled assembly we focus on two experimental observables, the fraction of subunits in capsids $\fc$ and the packaging efficiency, $\fp$, meaning the fraction of cores  contained in complete capsids.  To simplify the analysis we assume cooperative association of capsid proteins to cores, meaning that we neglect partially assembled intermediates. A full analysis including intermediates is performed in Ref. \cite{Hagan2009}. We first consider $\rho<\rhoStar$ so no empty capsid assembly occurs and $\fc= r \fp$. The dependence of $\fc$ and $\fp$  on the system parameters can be understood from three asymptotic limits, depending on the stoichiometric ratio $r$, which were described by Zandi and van der Schoot \cite{Zandi2009}. We then consider a fourth limit, $\rho>\rhoStar$.

\begin{enumerate}
\item $r\ll1$ and $\rho<\rhoStar$. For sufficient excess subunit, the free subunit concentration can be treated as approximately constant. Using Eq.~\ref{eq:Zcore} and following the analysis leading to equation Eq.~\ref{eq:fcAsympt} obtains
    \begin{align}
      \fc = &\approx r \frac{\left(\rho/\rhoStarStar \right)^N}{1+\left(\rho/\rhoStarStar \right)^N}  \\
      \rhoStarStar v_0 & = \exp\left[\beta \Gcore_N/N\right]
\label{eq:fcLowR},
\end{align}
where the threshold concentration $\rhoStarStar$ is smaller than that for empty capsids $\rhoStar$ due to the subunit-core interactions.

\item $r=1$ and $\rho<\rhoStar$. For exactly enough  capsid protein to encapsidate every core the concentration of free cores is related to that  of free capsid protein by $x_0=\rho_1/N$ and we obtain the same form as for empty capsids
\begin{align}
\fc & \approx \left(\frac{\rhoTot}{\rhoStar}\right)^N \ll 1 \qquad & \mbox{ for } \rhoTot < \rhoStarStar \nonumber \\
& \approx 1- \frac{\rhoStarStar}{\rhoTot} & \mbox{ for } \rhoTot > \rhoStarStar
\label{eq:fcRone}
\end{align}
but with $\rhoStarStar$ as in Eq.~\ref{eq:fcLowR}.

\item $r\gg1$ and $\rho<\rhoStar$. For excess core and cooperative subunit-core complexation (since we are neglecting partially covered cores) we obtain Eq. \ref{eq:fcRone} but with
\begin{align}
\rhoStarStar  = \exp\left(\beta\Gcore_N/N\right)\left(N \coreTot \right)^{-1/N}
\label{eq:fcHighR}
\end{align}
where we have assumed $N\gg1$.

\item $\rho>\rhoStar$. The conditions above apply in the limit $\rho<\rhoStar$ where empty capsids cannot assemble. For larger subunit concentrations empty capsids compete with assembly on cores. Due to the entropy cost of assembling a capsid on a core, Eq.~\ref{eq:Zcore} shows that there is a threshold surface free energy, $\Gcore_{N,0}-\Gcap_N \approx -k_{\text{B}}T \log (\CC a^3)$ for a stoichiometric amount of capsid protein and nanoparticles, above which full capsids are favored over empty capsids and free cores \cite{Hagan2009}. However, if nucleation of empty capsids ooccurs with a rate comparable to that on cores, complete or partially assembled empty capsids can assemble as a kinetic trap.

\end{enumerate}

Zandi and van der Schoot also investigated the effect of stoichiometric ratio on the size of the assembled capsid. They found that for large $r$ (i.e. excess genome) smaller capsids form, consistent with experimental observations \cite{Verduin1969,Sikkema2007}.

\subsection{Single-stranded RNA (ssRNA) encapsidation}
\label{sec:cargoDrivingForces}
As noted in the introduction, most ssRNA capsids assemble around their genome during replication. This process is driven to a large extent by electrostatic interactions between positive charges on the capsid proteins and negative charges on the RNA molecule. For example, the capsid proteins of many negative-stranded RNA viruses bind RNA via a positively-charged cleft \cite{Ruigrok2011}. For many positive-stranded RNA viruses, the capsid proteins bind RNA via a terminal tail rich in basic amino acids, which extends into the center of the virion and is typically unresolved in crystallographic maps\cite{Tong1993,Smith2000,Silva1985,Fisher1993,Choi1997,Speir1995}. These peptide tails are commonly referred to as arginine rich motifs (ARMs).

A remarkable set of \emph{in vitro} experiments indicates that just the negative charge found on the RNA phosphate groups is sufficient  to drive ssRNA capsid assembly around a cargo.  Namely, experiments have been performed in which  capsid proteins assemble into icosahedral capsids around various cargoes including genomic or non-genomic RNA (e.g. \cite{Bancroft1968,Bancroft1969,Hohn1969,Bancroft1970,Hiebert1968,Verduin1969,Krol1999,Kler2012,Cadena-Nava2012,Hu2008d}),  synthetic polyelectrolytes \cite{Hiebert1968,Bancroft1969,Sikkema2007,Hu2008,Comellas-Aragones2009,Brasch2012,Brasch2011,Kostiainen2011,Ma2012}, charge-functionalized nanoparticles  \cite{Dixit2006,Loo2006,Goicochea2007,Huang2007,Loo2007,Sun2007,Tsvetkova2012}, DNA micelles \cite{Kwak2010} and nano-emulsions \cite{Chang2008}.
However, as discussed further below features specific to biological RNA molecules such as their tertiary structure and sequence-specific interactions  likely promote selective assembly around the viral genome or help direct assembly toward particular morphologies.

  {\bf Optimal genome length.}
  The importance of nonspecific electrostatic interactions in driving RNA packaging has  been proposed as a constraint on the length of viral genome. For instance, Belyi and Muthukumar \cite{Belyi2006} and Hu and Shklovskii \cite{Hu2008a} reported a striking correlation between the total number of positive charges in the tails and the length of the genomic RNA. Interestingly, most ssRNA viruses are overcharged, meaning that the charge on the encapsidated RNA exceeds that of the charge on the inner surface of the capsid, often by about a factor of two. In support of this proposal, experiments on various viruses showed that partial deletions of the positively-charged residues in ARMs lead to virions that packaged a reduced amount of viral RNAs as compared to the wild-type capsid proteins \cite{Dong1998,Kaplan1998,Venter2009}.
%{\bf Optimal length of an encapsidated polyelectrolyte.}
 In further support of this proposal, a number of theoretical works have investigated the free energy to encapsidate a linear polyelectrolyte as a function of its length. The primary quantity of interest has been the thermodynamic optimal charge ratio, or the ratio between the negative charge on the polyelectrolyte and the positive charge on the capsid surface or peptide tails (ARMs) that minimizes the free energy.  It has been proposed that the observed correlation reflects the functional dependence of the optimum charge ratio, thus  suggesting  a thermodynamic constraint on the co-evolution of genome length and capsid charge. We summarize the results of these calculations here.

 Hu and Shklovskii assumed that RNA wraps around peptide tails and found that the free energy is minimized when the RNA contour length is approximately equal to the total contour length of peptide tails, resulting in an optimal charge ratio of packaged nucleotides to capsid charges of $2:1$.
Belyi and Muthukumar \cite{Belyi2006} treated the RNA as a polyelectrolyte and the peptide tails as oppositely charged brushes and used the ground state dominance approximation to predict an optimal charge ratio of $1:1$.  They noted that if the charge on the RNA and the peptide tails were renormalized according to counterion condensation theory \cite{Manning1969b}, the predicted ratio of packaged nucleotides to capsid charges would be $1.6:1$.  However, it is worth noting that condensed counterions are released by RNA-peptide association and thus factor into the encapsidation free energy. Our simulations with explicit ions and polymers with different linear charge densities (unpublished) suggest that the bare charge density should be used. Calculations that placed the capsid charge entirely at the surface predicted charge ratios of $\lesssim 1:1$ \cite{Siber2008} and $2:1$ \cite{Schoot2005}. Siber et al. showed that the optimal charge ratio was less dependent on ionic strength when the capsid charge moved off of the inner surface, but still found an optimal charge ratio of $<1:1$. The dependence of the encapsidation free energy on the length of a linear polyelectrolyte has also been investigated in the limit of no added salt using Monte Carlo simulations on a model which assumes neither the continuum limit nor spherical symmetry \cite{Angelescu2006}. These calculations also predicted that the optimal genome length would correspond to a charge ratio of $1:1$.

 Ting et al. \cite{Ting2011} used a self consistent field theory to show that the optimal ratio of packaged charge to capsid charge depends sensitively on excluded volume and thus varies with many factors including capsid size and charge density on peptide arms. Their model predicted an optimal charge ratio of less than $1:1$ for all relevant parameters. They noted however that the cytoplasm contains a significant concentration of negative charge in the form of large macromolecules which are excluded from the interior of an assembled capsid. They found that the Donnan potential due to this excluded charge is essential to obtain a thermodynamic optimum charge ratio which is larger than 1:1, and thus suggested that this effect plays an important role in capsid assembly \emph{in vivo}. However, recent \emph{in vitro} competition assays in which different RNAs compete for packaging under conditions of limiting capsid protein showed that longer RNAs (up to viral genome lengths) were preferentially packaged over shorter RNAs by CCMV capsid proteins \cite{Comas-Garcia2012}. This result suggests that the genome length is nearly optimal for packaging even in the absence of a Donnan potential.
 An interesting and somewhat surprising prediction of the self consistent field theory \cite{Ting2011} is that the encapsidation free energy was essentially the same for a linear polyelectrolyte and for a polyelectrolyte with a branched structure reflecting the secondary and tertiary structure of an RNA molecule.

Although the quantitative results of these theoretical calculations vary depending on their assumptions, they agree with experiments that the optimal genome length is an increasing function of the capsid charge. They cannot explain the observation that polymers with apparent charge ratios as large as 9:1 were encapsidated in Ref. \cite{Hu2008d}; however, the predictions above are for the optimal length and it was not possible to measure the packaging efficiencies in those experiments. Models based on linear polyelectrolytes also do not capture the sequence dependence of the charge ratios observed in recent \emph{in vivo} experiments for mutant BMV capsid proteins \cite{Ni2012} (discussed next).
This discrepancy suggests that sequence-specific RNA-protein interactions which have not been accounted for in these calculations play a role in RNA packaging during assembly.

   Several sets of experiments indicate sequence-specific roles for the RNA and proteins.  Recently, Ni and coworkers \cite{Ni2012} performed an extensive investigation of virions assembled in \emph{N. benthamiana} plants from BMV proteins with positively-charged residues added, deleted, or substituted. They found a correlation between the amount of positive charge on the capsid and the amount of packaged RNA. However, relationships between mutations and the amounts and sequences of packaged RNA were sensitive to factors other than charge, indicating that  that the charge on the capsid is not the sole factor in determining the amount or types of RNA which are packaged \emph{in vivo}. \emph{In vitro} experiments have also shown that the BMV capsid preferentially encapsidates RNAs containing a tRNA-like structure \cite{Choi2002}. In cells, interactions between the capsid protein and this tRNA-like structure may play a role in packaging specificity by coordinating RNA replication and encapsidation \cite{Annamalai2006,Annamalai2005,Annamalai2008}. Packaging signals, or regions of RNA that have sequence-specific interactions with the capsid protein, are known for some viruses (e.g. HIV \cite{Pappalardo1998,Lu2011,D'Souza2005}) or MS2 and satellite tobacco necrosis virus (STNV) \cite{Bunka2011,Borodavka2012}). As discussed in section \ref{sec:largeTDynamics} the MS2 RNA regulates conformational switching in a sequence-dependent manner \cite{Basnak2010}.

{\bf Genome organization.}
 A number of studies have simulated bead-spring models of polyelectrolytes confined within simplified representations of viral capsids. In some cases the capsid was modeled by a spherical container with different arrangements of embedded charges \cite{Angelescu2007,Angelescu2006}, while others were based on particular capsid structures.
 For example, Zhang et al. \cite{Zhang2004} incorporated the electrostatic potential derived from the CCMV while the model of Forrey and Muthukumar \cite{Forrey2009} captured the charge distribution of  the inner surface of the Pariacoto virus crystal structure. These studies showed that  a dodecahedral arrangement of surface charges \cite{Angelescu2007} or basic charges on peptide tails \cite{Forrey2009} leads the encapsulated polymer to adopt a dodecahedral cage structure. A model of capsid assembly dynamics \cite{Elrad2010} also found that the polymer adopted the symmetry of the overlying capsid charge.
 ElSawy et al. \cite{ElSawy2011} showed that a bead-spring polyelectrolyte placed inside an atomic model of MS2  could  reproduce the observed multiple peaks of the radial RNA density distribution \cite{Toropova2008} and the icosahedral order of the outer layer. Their simulations suggested that RNA-RNA repulsion and the arrangements of crevices on the inner capsid surface were the dominant forces directing organization of the outer layer of genome.
Several models have used the detailed knowledge of RNA density \cite{Dykeman2011} and its icosahedral order \cite{Dykeman2011,Rudnick2005,Bruinsma2006,Stockley2007} to suggest RNA-directed assembly pathways.

 {\bf Modeling assembly around nanoparticles.} Hagan \cite{Hagan2009} and Zandi et al. \cite{Siber2010} used self consistent field theories to calculate the interaction between positive charges on capsid protein tails and carboxyl groups functionalized on the surface of nanoparticles \cite{Chen2006,Chen2005,Daniel2010,Goicochea2007,Huang2007,Sun2007,Tsvetkova2012}, enabling prediction of time-dependent \cite{Hagan2009} or equilibrium packaging efficiencies \cite{Hagan2009,Siber2010}. Siber et al. \cite{Siber2010} also considered assembly into different icosahedral morphologies. In Ref. \cite{Hagan2009} the kinetics of assembly around corners were also considered by estimating the free energy $\Gcore_n$ for cores partially covered with intermediates and extending the rate equation approach described in section \ref{sec:rateEquations} to describe the kinetics of core-controlled assembly. Both the kinetic and equilibrium approaches predicted a threshold density of functionalized charge below which no significant assembly on cores would take place. This result was qualitatively confirmed by subsequent experiments  \cite{Daniel2010}. Siber \cite{Siber2010} et al also predicted that high charge functionalization densities and  excess capsid protein would favor pseudo-$T$=2 capsids, which was seen experimentally by Daniel et al. \cite{Daniel2010}.

\subsection{Dynamics of assembly around cores}
\label{sec:cargoDynamics}
{\bf Nanoparticles.}
 As noted in the previous section Hagan \cite{Hagan2009} used polyelectrolyte theory to calculate the electrostatic driving force for capsid proteins to adsorb onto nanoparticle surfaces as a function of the  nanoparticle surface charge density. This dependence was used to extend the rate equation approach (section \ref{sec:rateEquations}) to predict assembly kinetics around attractive cores. The calculations found a threshold density of functionalized charge, above which capsids efficiently assemble around nanoparticles, and that light scattering increases rapidly at early times, without the lag phase characteristic of empty capsid assembly. These results were consistent with experimental measurements.

{\bf Polymers.}
 Elrad and Hagan \cite{Elrad2008} developed a coarse-grained computational model that describes the assembly dynamics of icosahedral capsids from subunits that interconvert between different conformations (section \ref{sec:largeTDynamics}).  The simulations identified mechanisms by which subunits form empty capsids with only one morphology, but adaptively assemble into different icosahedral morphologies around nanoparticle cargoes with varying sizes, as seen in experiments \cite{Sun2007}. Adaptive cargo encapsidation required moderate cargo-subunit interaction strengths; stronger interactions frustrated assembly by stabilizing intermediates with incommensurate curvature.

 Kivenson and Hagan \cite{Kivenson2010} explored capsid assembly around a flexible polymer with a model defined on a cubic lattice using dynamic Monte Carlo, which allowed simulation of large capsid-like cuboidal shells over long time scales. By simulating assembly with a wide range of capsid sizes and polymer lengths, the simulations showed that there is an optimal polymer length which maximizes encapsulation yields at finite observation times. The optimal length scaled with the number of attractive sites on the capsid in the absence of attractive interactions between polymer segments. However, introducing attractive interactions between polymer segments, which physically could arise from base pairing or multivalent counterions, increased the predicted optimal length dramatically. A limitation of these simulations was that the Monte Carlo move set did not enable simultaneous motions of polymer segments and capsid subunits, and thus  could not accurately describe the dynamics for parameter sets for which cooperative polymer-subunit motions play an important role in assembly.

\begin{figure}
\begin{center}
\epsfxsize=0.6\columnwidth\epsfbox{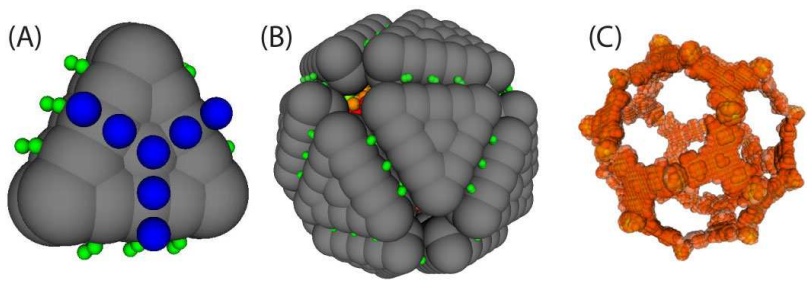}
\caption{The capsid model from Ref.\cite{Elrad2010}. {\bf (A)} The model subunit, as viewed from inside the capsid. The gray overlapping spheres interact via repulsive potentials \cite{Weeks1971}, complementary capsomer-capsomer attractors (green spheres at the subunit edges) experience attractive interactions and the capsomer–-polymer attractors (blue spheres on the subunit inner surface) experience short-range attractions to polymer segments.  Sphere sizes indicate the interaction length scale.  {\bf (B)} Image of a well-formed capsid assembled around a polymer (shown in red). {\bf (C)} Visualization of the polymer density inside the capsid. The polymer density is averaged over a large number of successful assembly trajectories
after completion, for a polymer with length $N_\text{p} = 150$ segments. Densities are averaged over the threefold symmetry of the capsomer, but not over the
20-fold symmetry group of the completed capsid. Images reprinted with permission from Phys. Biol., {\bf 7}, 045003 (2010), Elrad and Hagan, \emph{Encapsulation of a polymer by an icosahedral virus}, Copyright (2010) IOP Publishing. }
\label{fig:Elrad-model}
\end{center}
\end{figure}
\begin{figure}
\begin{center}
\epsfxsize=0.42\textwidth\epsfbox{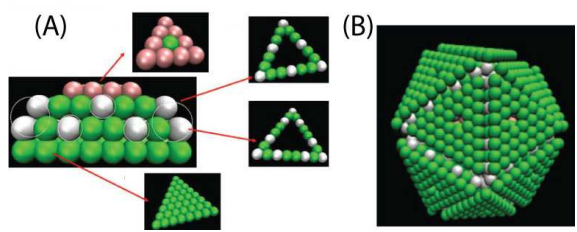}
\caption{The capsid model from Ref. \cite{Mahalik2012}. {\bf (A)} The model subunit.  All beads experience repulsive excluded-volume interactions. In addition, pairs of white beads  experience short-range attractive interactions and the pink beads have a charge of $+e$. Electrostatics interactions are represented by Debye Huckel interactions. {\bf (B)}. The low-energy capsid structure. Images reprinted with permission from J. Chem. Phys., {\bf 136}, 135101 (2012), \emph{Langevin dynamics simulation of polymer-assisted virus-like assembly}, Mahalik and Muthukumar, Copyright(2012) American Institute of Physics.}
\label{fig:Mahalik-model}
\end{center}
\end{figure}

 Elrad and Hagan \cite{Elrad2010} performed Brownian dynamics simulations of encapsidation of a flexible polymer by a model capsid with icosahedral symmetry (Fig. \ref{fig:Elrad-model}). The model considered a truncated-pyramidal extended subunit model of the form used to the assembly of  empty $T$=1 capsids from trimeric subunits \cite{Rapaport1999,Rapaport2004,Rapaport2008,Hagan2011} (section \ref{sec:particleBased}). The polymer was represented by a flexible bead-spring model, with repulsive interactions between segments corresponding to screened electrostatic repulsions, and short-range attractions to `charge sites' located on one surface of the model subunits, corresponding to screened attractive electrostatic interactions between opposite charges on the polymer and capsid subunits.  Use of the short-range interactions (which increased computational efficiency) was justified by the fact that capsid assembly \emph{in vivo} or \emph{in vitro} always occurs  at moderate salt concentrations for which the Debye screening length is on the order of 1 nm.

 Subsequently Mahalik and Muthukumar \cite{Mahalik2012} simulated the assembly of extended triangular subunits (Fig.~\ref{fig:Mahalik-model}) around a linear polyelectrolyte with  screened electrostatics modeled by Debye Huckel interactions. The simulations led to predictions which were qualitatively similar to those of Ref. \cite{Elrad2010}, but the  Debye Huckel interactions do allow for a more straightforward connection to experimental salt concentrations.

\begin{figure}[bt]
\begin{center}
\epsfxsize=0.72\textwidth\epsfbox{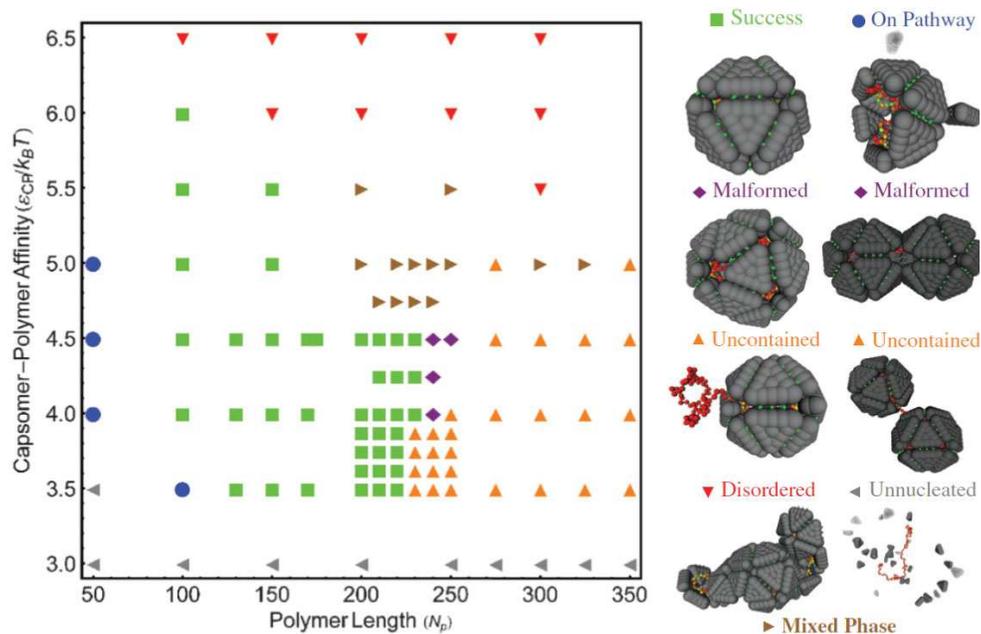}
\caption{Kinetic phase diagram showing the dominant assembly product as a function of polymer length $\lp$ and capsomer-polymer interaction strength $\ecp$ for capsomer-polymer interaction strength $\ecc = 4.0 \kt$ and subunit concentration $\logc = -7.38$. The legend on the right shows snapshots from simulations that typify each dominant configuration. Figure reprinted with permission from Phys. Biol., {\bf 7}, 045003 (2010), Elrad and Hagan, \emph{Encapsulation of a polymer by an icosahedral virus}, Copyright (2010) IOP Publishing}
\label{fig:Elrad-phase-diagram}
\end{center}
\end{figure}

\begin{figure}
\begin{center}
\epsfxsize=0.2\textwidth\epsfbox{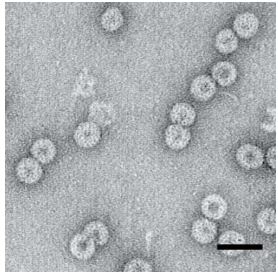}
\caption{Doublet virus-like particles assembled from CCMV capsid proteins around tobacco mosaic virus (TMV) RNA, with 6400 nucleotides or approximately twice the number of nucleotides packaged in a native CCMV virion \cite{Cadena-Nava2012}. Negative-stain transmission electron microscopy images are shown and the scale bar is 50 nm. Image provided by C. Knobler and W. Gelbart.
%Image from Ref. \cite{Cadena-Nava2012}.
}
\label{fig:ExperimentalMultiplets}
\end{center}
\end{figure}

{\bf Assembly outcomes.}
Simulations were performed over wide ranges of polymer lengths, subunit concentrations, subunit-polymer interaction strengths, and subunit-subunit interactions strengths. Refs. \cite{Kivenson2010,Elrad2010,Mahalik2012} found that productive assembly around the polymer can occur for protein-protein interaction strengths and protein concentrations for which empty capsid assembly does not occur, as observed for many ssRNA viruses. The polymer-protein interactions stabilize protein-protein interactions, lowering the nucleation barrier and enhancing the thermodynamic favorability of the assembled capsid.
Ref. \cite{Elrad2010} presented
the assembly yields and assembly morphologies at long (but finite) times in a phase diagram. These observables  can be experimentally characterized using EM.
As shown in Fig. \ref{fig:Elrad-phase-diagram}, there is a region of parameter space in which  assembly is `successful' meaning that the predominate assembly product is a well-formed capsid which completely encapsulates the polymer. For stronger than optimal subunit-subunit or subunit-polymer interaction strengths the system becomes trapped in metastable disordered states. Longer than optimal polymer lengths can give rise to closed but defective capsids or  partially complete capsids. Polymer lengths that were significantly larger than optimal gave rise to structures in which multiple nearly complete capsids were connected by an RNA molecule.  Polymers on the order of twice the optimal length gave rise to doublet capsids (Fig. \ref{fig:Elrad-phase-diagram} right), while longer polymers gave rise to higher order multiplet capsids.

Interestingly, structures with similar morphologies were observed in \emph{in vitro} experiments in which CCMV capsid proteins assembled around RNA molecules with lengths in excess of the viral genome length \cite{Cadena-Nava2012} or conjugated polyelectrolytes  \cite{Brasch2012}. For example, Fig. \ref{fig:ExperimentalMultiplets} shows virus-like particles assembled around TMV RNA  with 6400 nucleotides (nt), or approximately twice the number of nucleotides of the native CCMV RNA typically encapsulated within a single virion. These images can be compared to the right-most image labeled `uncontained' in Fig. \ref{fig:Elrad-phase-diagram}.   The fact that the capsids were connected by RNA was confirmed by showing that the multiplets separated upon introduction of RNAase. RNA lengths of 9000 or 12000 nt respectively led to triplet and quadruplet capsids \cite{Cadena-Nava2012}.

\begin{figure}
\begin{center}
\includegraphics[width=0.6\textwidth]{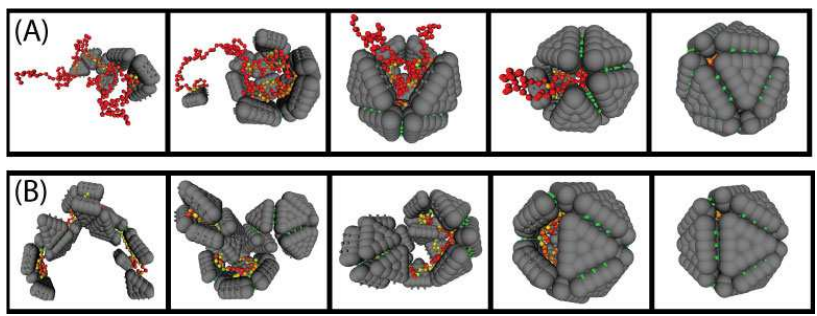}
\caption{Two mechanisms for assembly around a polymer \cite{Elrad2010}. {\bf (A)} Strong subunit-subunit interactions and relatively weak subunit-polymer interactions led to a nucleation and growth mechanism, where first a small partial capsid formed on the polymer followed by sequential addition of subunits.  {\bf (B)} Weaker subunit-subunit interactions and stronger subunit-polymer interactions led to a disordered assembly mechanism, where more than 20 subunits (the size of a complete capsid) bound to the polymer in a disordered arrangement,  followed by annealing of multiple intermediates and finally completion. Figure reprinted with permission from Phys. Biol., {\bf 7}, 045003 (2010), Elrad and Hagan, \emph{Encapsulation of a polymer by an icosahedral virus}, Copyright (2010) IOP Publishing.}
\label{fig:Elrad-assembly-mechanism}
\end{center}
\end{figure}

\begin{figure}
\begin{center}
\includegraphics[width=0.7\columnwidth]{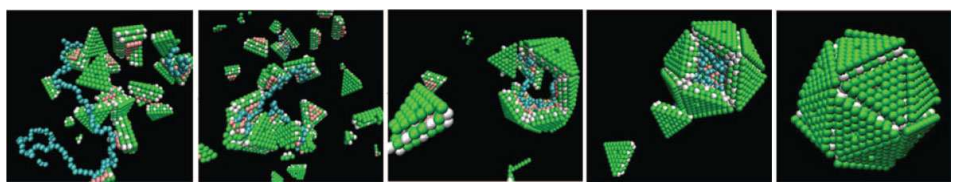}
\caption{Snapshots from a simulation in Ref. \cite{Mahalik2012}, in which assembly proceeded by the  disordered assembly mechanism. Images reprinted with permission from J. Chem. Phys., {\bf 136}, 135101 (2012), \emph{Langevin dynamics simulation of polymer-assisted virus-like assembly}, Mahalik and Muthukumar, Copyright(2012) American Institute of Physics.  }
\label{fig:Mahalik-assembly-mechanism}
\end{center}
\end{figure}

{\bf Assembly mechanisms.}
The simulations in Ref. \cite{Elrad2010} also demonstrated that there are two classes of assembly mechanisms that can occur around a central core (i.e. RNA or a polymer), each of which leads to a different assembly kinetics (Fig. \ref{fig:Elrad-assembly-mechanism}). One  closely resembles the nucleation and growth mechanism by which empty capsids assemble, except that the polymer plays an active role by stabilizing protein-protein interactions and by enhancing the flux of proteins to the assembling capsid (Fig. \ref{fig:Elrad-assembly-mechanism}A). A small partial capsid first nucleates on the polymer, followed by a growth phase in which one or a few subunits sequentially and reversibly add to the partial capsid. Polymer encapsulation proceeds in concert with capsid assembly. In the alternative mechanism, first proposed by McPherson \cite{McPherson2005}, subunits adsorb onto the polymer en masse in a disordered fashion and then cooperatively rearrange to form  an ordered capsid (Fig. \ref{fig:Elrad-assembly-mechanism}B). In many cases excess subunits adsorb onto the polymer, so that subunit dissociation is required for assembly to complete. A similar mechanism was observed by Mahalik and Muthukamar \cite{Mahalik2012} (Fig. \ref{fig:Mahalik-assembly-mechanism}).

The simulation results predict that the assembly mechanism can be tuned experimentally by changing charge densities, solution conditions, or assembly protocols. The nucleation and growth mechanism is favored when protein-polymer association is weak or slow (e.g. at high salt concentration) and protein-protein interactions are strong. The disordered assembly mechanism arises when protein-polymer association is strong and rapid, so that complete coverage of the polymer occurs before significant rearrangements of already adsorbed proteins can lead to assembly. In support of this idea, the \emph{in vitro} kinetics for CCMV described in Ref. \cite{Johnson2004} appear consistent with the nucleation and growth mechanism, whereas the assembly protocol used by Cadena-Nava and coworkers \cite{Cadena-Nava2012} favored the disordered assembly mechanism.

{\bf Assembly kinetics.}
The two mechanisms described above give rise to very different dependencies of assembly kinetics on control parameters such as protein concentration or interaction strengths. In the case of the nucleation and growth mechanism, the assembly time can be described in terms of the timescale for each phase, $\tnuc$ and $\telong$ respectively (Eq. \ref{eq:assemblyTime}) as for empty capsids. The assembly kinetics depend on the stoichiometric ratio of total polymer concentration $\rhopt$ to capsid protein concentration $\rhoTot$, $\stoich = N \rhopt/\rhoTot$, with different relationships in the limits of excess protein, excess polymer, or stoichiometric amounts of protein in polymer.

For brevity we analyze only one case here, the case of excess capsid protein, $\stoich\gg1$. The other stoichiometric limits can be understood via similar arguments. In this case the capsid protein concentration remains nearly constant during  the course of assembly. Recall (section \ref{sec:assemblyTimeScales}, \eqref{eq:tnuc}) that concentrations of partial capsids which are smaller than the critical nucleus size $\nnuc$ can be treated as in quasi-equilibrium, and the nucleation rate is given by the rate at which subunits associate with the largest pre-nucleus, which has $\nnuc-1$ subunits. In the limit $\stoich\gg1$ the concentration of pre-nuclei adsorbed on polymers can be expressed as
\begin{equation}
\rho_{\nn}\cong \rhop \Lp (\rho_1 v_0)^{\nn} \exp(-\beta G_\nn)
 \label{eq:polymer}
 \end{equation}
 with $G_\nn$ the interaction energy for the $\nn$-sized complex, including  protein-polymer interactions. The latter were found to be crucial in both Ref. \cite{Elrad2010} and Ref. \cite{Mahalik2012}, as they effectively enhance the protein concentration in the vicinity of the polymer and thus lower the nucleation barrier.  The factor $\rhop \Lp$, with $\rhop$ the concentration of polymers which are not covered by capsids and $\Lp$ the polymer length accounts for the fact that the concentration of pre-nuclei adsorbed on polymers must be proportional to the number of available adsorption sites and hence is proportional to the total concentration of segments on uncovered polymers. The nucleation rate is then given by the rate of one additional subunit adding to the pre-nucleus complex, $f \rho_1$ where $f$ is a rate constant, to yield
\begin{align}
-\frac{d \rhop}{dt} &= \rhop \tnuc \nonumber \\
\tnuc &\cong f \Lp (v_0 \rho_1)^{\nnuc}\exp(-\beta G_\nn)
 \label{eq:polymerNuc}.
 \end{align}
 This equation is first order in polymer concentration, and the timescale for depleting polymers is given by $\tnuc$. The inverse dependence of $\tnuc$ on polymer length was observed in Ref. \cite{Kivenson2010}.

\begin{figure}
\begin{center}
\includegraphics[width=0.5\textwidth]{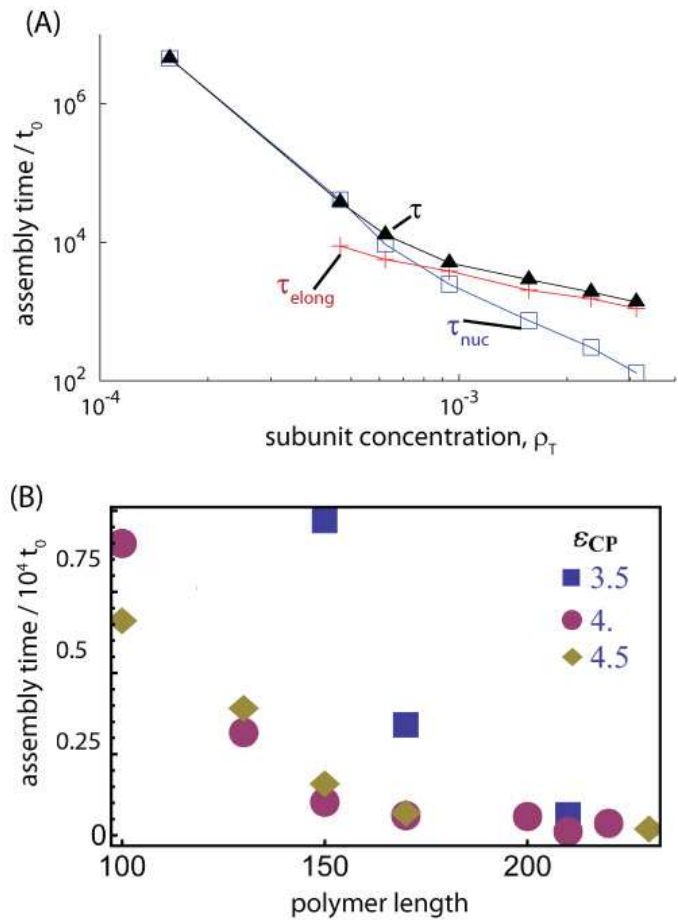}
\caption{Assembly times depend on subunit concentration and polymer length in simulations of assembly around a polymer. {\bf (A)} The median overall assembly time $\tau$ ($\blacktriangle$ symbol), nucleation time $\tnuc$ ($\Box$ symbol), and elongation time $\telong$ ($+$ symbol) are shown as a function of subunit density for the model in Fig.~\ref{fig:Elrad-model}. Simulations at the lowest concentration  used forward flux sampling \cite{Allen2006,Allen2005} to overcome the large nucleation barrier. {\bf (B)} The median elongation time is shown as a function of polymer length for several values of the polymer-subunit interaction strength, $\ecp$. The plot in (B) is reprinted with permission from Phys. Biol., {\bf 7}, 045003 (2010), Elrad and Hagan, \emph{Encapsulation of a polymer by an icosahedral virus}, Copyright (2010) IOP Publishing. }
\label{fig:polymerAssemblyTimes}
\end{center}
\end{figure}

As in the case of empty capsids, there is a second timescale $\telong$ describing  the time required for a nucleated partial capsid to grow to completion. Hu and Shklovskii \cite{Hu2007} proposed that the polymer could substantially decrease the elongation time in comparison to the case of empty capsid assembly because the polymer acts as an `antenna' which increases the  collisional cross-section. In analogy to transcription factors searching for their binding sites on DNA \cite{Berg1981,Hu2006} adsorbed subunits can undergo one-dimensional diffusion, or sliding, on the polymer. It was predicted that the elongation time would thus decrease with increasing polymer length (for fixed capsid size) \cite{Hu2007} as
\begin{equation}
\telong \sim \Lp^{-1/2} \rho_1^{-1}
\label{eq:telongPolymer}.
\end{equation}
Here we have assumed that subunits add independently during elongation and thus $\telong$ is inversely proportional to the subunit concentration  according to the arguments in section \ref{sec:assemblyTimeScales}. The simulations in Refs. \cite{Elrad2010} found that the polymer did significantly increase assembly rates (Fig. \ref{fig:polymerAssemblyTimes}A), through a combination of subunit sliding and cooperative subunit-polymer motions, where the polymer drags associated subunits to the partial capsid like the action of a fly-fishing line with a hooked fish (or the Cookie Monster bringing cookies to his mouth). Enhanced elongation rates were also observed by Mahalik and Muthukumar \cite{Mahalik2012}. However, neither study was able to consider a wide enough range of polymer lengths to investigate the scaling predicted by Eq.~\ref{eq:telongPolymer}.

Combining Eqs.~\ref{eq:polymerNuc} and \ref{eq:telongPolymer} gives the total assembly time $\tau = \tnuc + \telong$ as
\begin{equation}
\tau \sim \Lp^{-1} \rho_1^{-\nnuc} \exp(\beta G_\nn) + \Lp^{-1/2} \rho_1^{-1}
\label{eq:polymerTau}.
\end{equation}
Notice that the nucleation and elongation times have very different dependencies on the free subunit concentration $\rho_1$. Therefore, the concentration dependence of the overall assembly time $\tau$ depends on which of these processes is rate limiting. As shown in Fig. \ref{fig:polymerAssemblyTimes}A, median assembly times measured by simulation show a crossover from nucleation limited behavior at low concentrations to  elongation limited behavior at large concentrations. As anticipated, the elongation time scales as $\telong \sim \rho_1^{-1}$ whereas the nucleation time scales with a larger, concentration dependent power (see section \ref{sec:nucleationElongation}). Notice that the elongation-limited regime observed at high subunit concentrations would not occur for empty capsid assembly,  where elongation limiting behavior would lead to the free subunit starvation  kinetic trap (section \ref{sec:assemblyTimeScales}, Eq. \eqref{eq:ckt}). That trap is avoided when capsid protein is in excess over polymers. However, the trap can occur for  stoichiometric  amounts of polymer and protein or excess polymer.

The simulations in Ref. \cite{Mahalik2012} find trends which are qualitatively similar to those found in Ref. \cite{Elrad2008} although the kinetics are analyzed in Ref. \cite{Mahalik2012} according to the framework of polymer crystallization kinetics theory. In Ref. \cite{Mahalik2012} parallel tempering was used to calculate the free energy profile as a function of partial capsid size, which showed that  the presence of  the polymer dramatically reduced the nucleation barrier. The polymer also reduced elongation times.

\section{Outlook}
\label{sec:conclusions}
In this review we have tried to present a summary of the theoretical and computational methodologies that have been used to  model capsid assembly. We have presented examples in which modeling led to important insights which either explained existing experiments or suggested new ones. We have also tried to identify limitations in computational power, model construction, or available experimental technology which hinder the connection between theory and experiment. Clearly, as the powers of computers and the computational techniques performed on them increase, it will become possible to model the assembly process with increasing resolution and to account for features of capsid proteins and other virion components not accounted for by current models.

Advances in experimental technologies may provide the most important opportunities for progress in modeling. Recent experiments that visualize or otherwise monitor the formation of individual viruses  (e.g. \cite{Baumgaertel2012,Jouvenet2011,Zhou2011,Harms2011}) can provide information about distributions of capsid growth times and the rates at which individual subunits associate to or dissociate from partial capsids of different sizes. This information would provide additional constraints with which to refine or extend current models. At the same time, the structures of most intermediates on assembly pathways remain challenging to characterize, and thus models which can predict pathways provide an important complement to these new methods.

The modeling efforts covered in this review have primarily focused on \emph{in vitro} experimental systems for the obvious reasons of reduced complexity and increased control over system parameters offered by the test tube. Looking ahead, it will be important for models to include effects relevant to the environment of host organisms, such as molecular crowding, compartmentalization, and coupling between translation and assembly. However, quantitative data from \emph{in vivo} experiments is essential for building such models, estimating their parameters, and for testing model predictions. Ultimately, by combining complementary \emph{in vivo} experiments, controlled \emph{in vitro} experiments in which key parameters can be tuned, and modeling that can elucidate individual assembly pathways and the effect of individual parameters, we can identify the factors that confer robustness or sensitivity to the process of virus assembly.

%\acknowledgement
{\bf Acknowledgement.} This work was supported by the NIH through Award Number R01AI080791 from the National Institute Of Allergy And Infectious Diseases.  I gratefully acknowledge Charles Knobler, Teresa Ruiz-Herrero, Rob Jack, and Dennis Rapaport for critical reads of the manuscript. Computational resources for the simulations performed for Fig. \ref{fig:polymerAssemblyTimes}A were provided by the National Science Foundation through XSEDE computing resources and the Brandeis HPCC.

\bibliographystyle{plain}
\bibliography{all-references}
\end{document}